\documentclass[letterpaper,11pt]{article}
\usepackage{amsmath, amsthm, amssymb}    
\usepackage{bridges}			
\usepackage{graphicx}			
\usepackage[colorlinks=true, urlcolor=blue, citecolor=black, linkcolor=black]{hyperref}  
\usepackage{subcaption}			

\usepackage[bottom]{footmisc} 

\usepackage[table]{xcolor}

\title{Applying the Iterative Development Process: \\The Creation of {\em Fractal Emergence}}

\author{Christopher R.\ H.\ Hanusa\textsuperscript{1} and Eric Vergo\textsuperscript{2}
\vspace{10pt}\\
\textsuperscript{1}Department of Mathematics, Queens College, CUNY, Queens, NY, USA; chanusa@qc.cuny.edu\\
\textsuperscript{2}Department of Mathematics, Queens College, CUNY, Queens, NY, USA; ericvergo@gmail.com} \date{}		

\begin{document}

\maketitle

\thispagestyle{empty}

\begin{abstract}
The iterative development process is a framework used to design products and applications across a wide range of domains. It centers around building prototypes, testing them, and updating based on the test results. We discuss how we applied this technique to create {\em Fractal Emergence}, an interactive piece of mathematical art. 
\end{abstract}

\section*{Introduction}

\begin{figure}[b]
	\centering
 \raisebox{1.9in}{(a)}~
 	\includegraphics[height=2.1in]{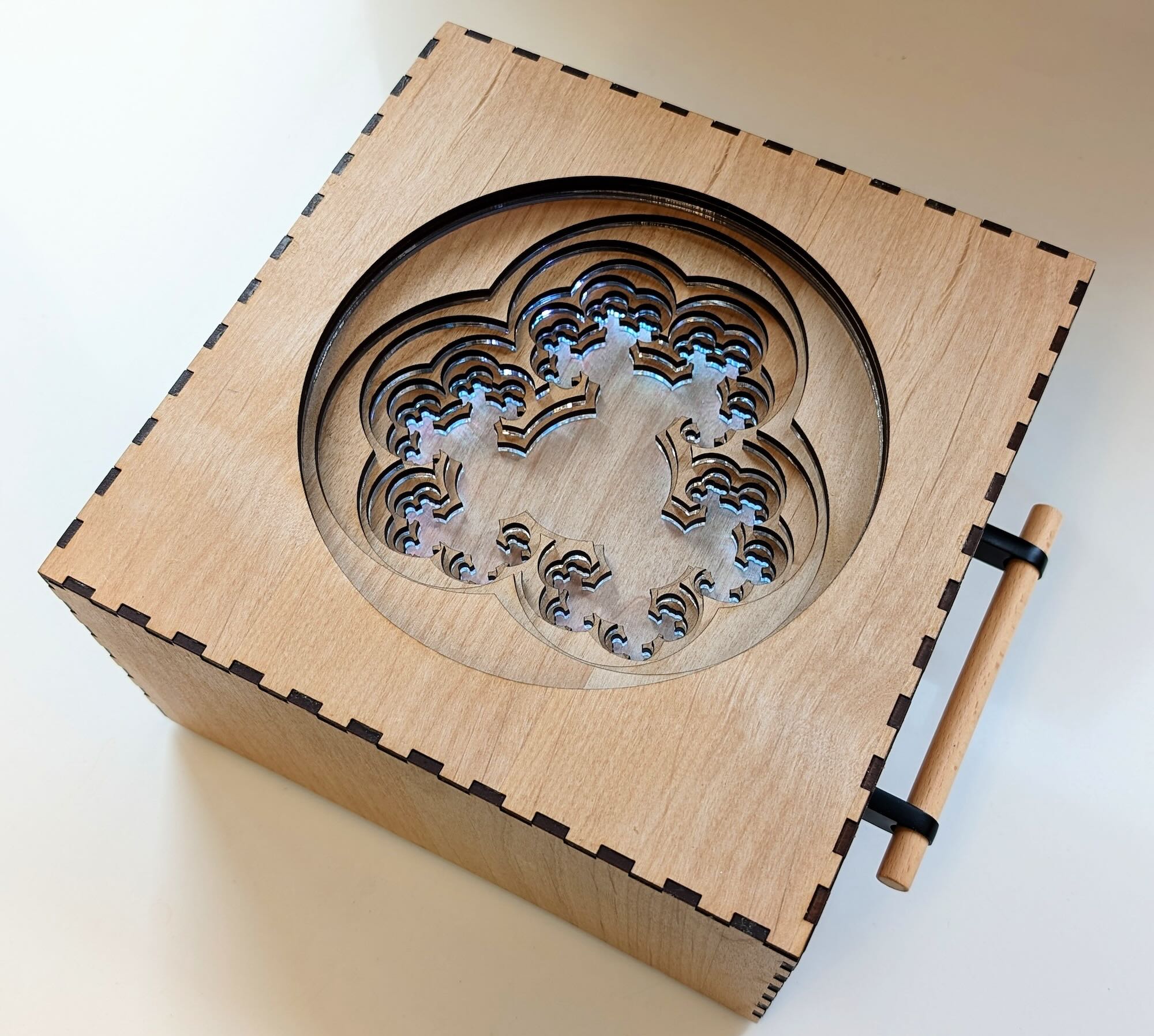}
    \qquad\qquad
 \raisebox{1.9in}{(b)}
    \includegraphics[height=2.1in]{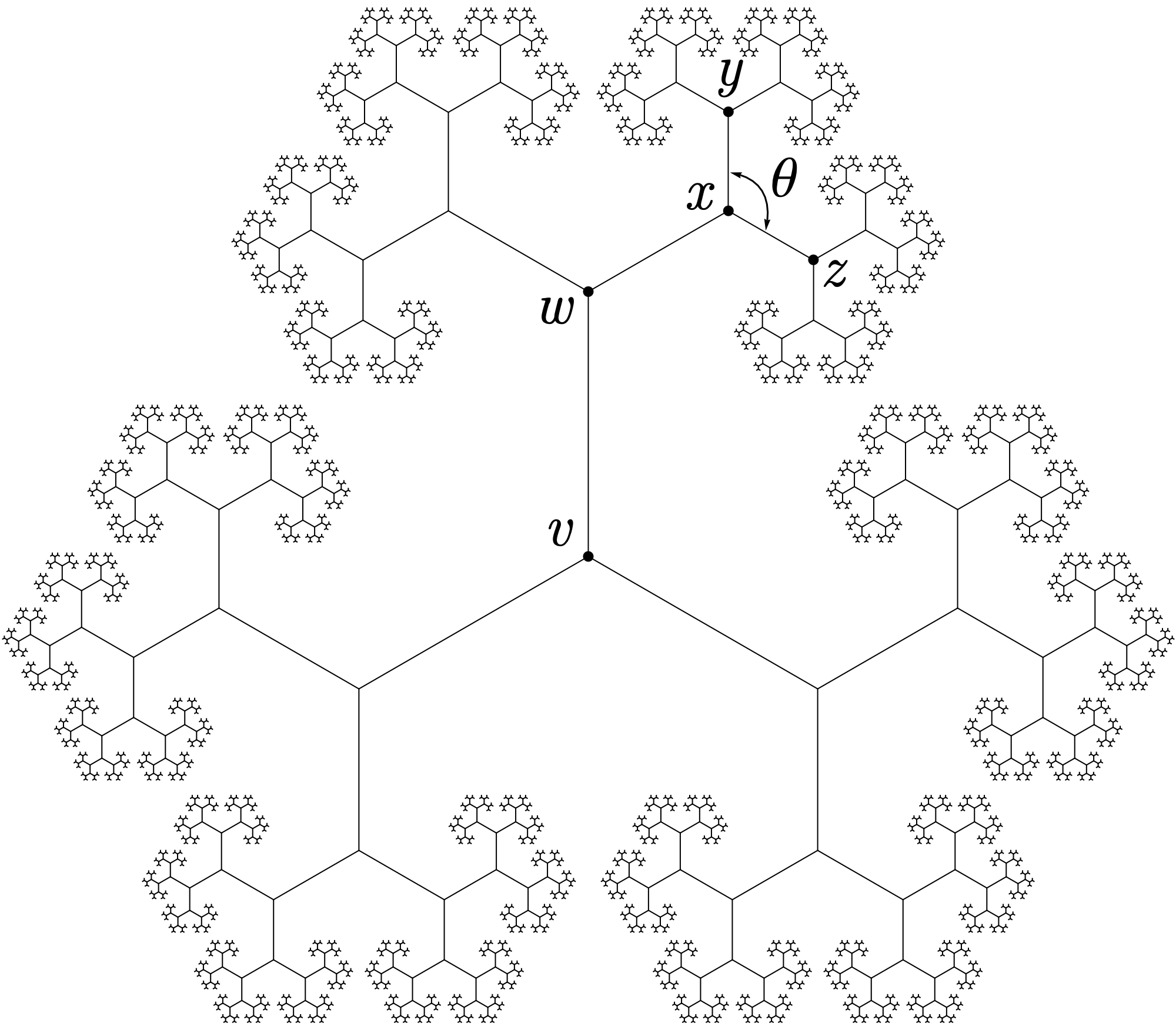}
 	\caption{(a) A still image of Fractal Emergence. (b) The infinite trivalent tree $\mathcal{T}$.}
  	\label{fig:fractalemergence}
 \end{figure}

This paper details the process through which the authors designed {\em Fractal Emergence}, an interactive light display that guides the viewer through the iterations that create a specific fractal based on the infinite trivalent tree. This piece is made from laser-cut wood and acrylic, with an internal mechanism that slides a hidden ring of LED lights up and down to cast light on layer after layer of the interior construction. See Figure~\ref{fig:fractalemergence}(a).

\begin{figure}[t]
 	\begin{tabular}{r}\includegraphics[height=1.6in]{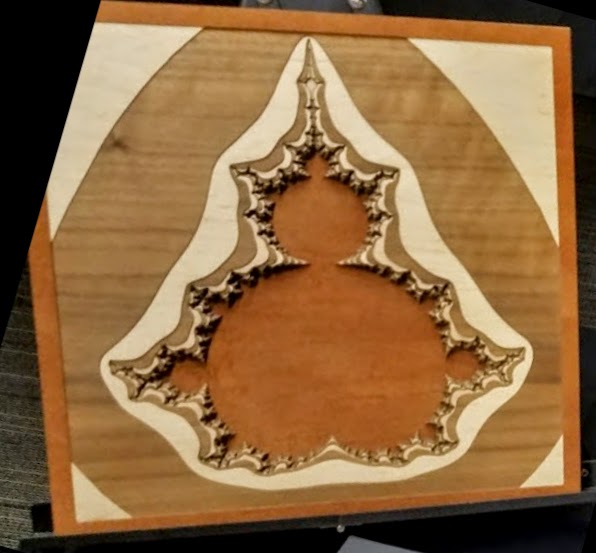}\vspace{-.05in}\\{\small photo by Christopher Hanusa}\end{tabular}\!\!\qquad\!\!%
    \begin{tabular}{r}\includegraphics[height=1.6in]{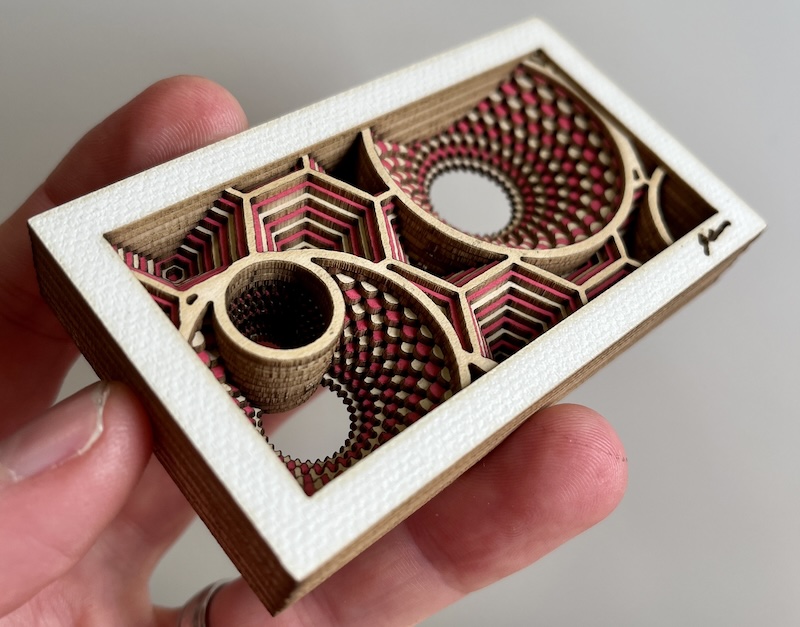}\vspace{-.05in}\\{\small photo by Shawn Kemp}\end{tabular}\!\!\qquad\!\!%
    \begin{tabular}{r}\includegraphics[height=1.6in]{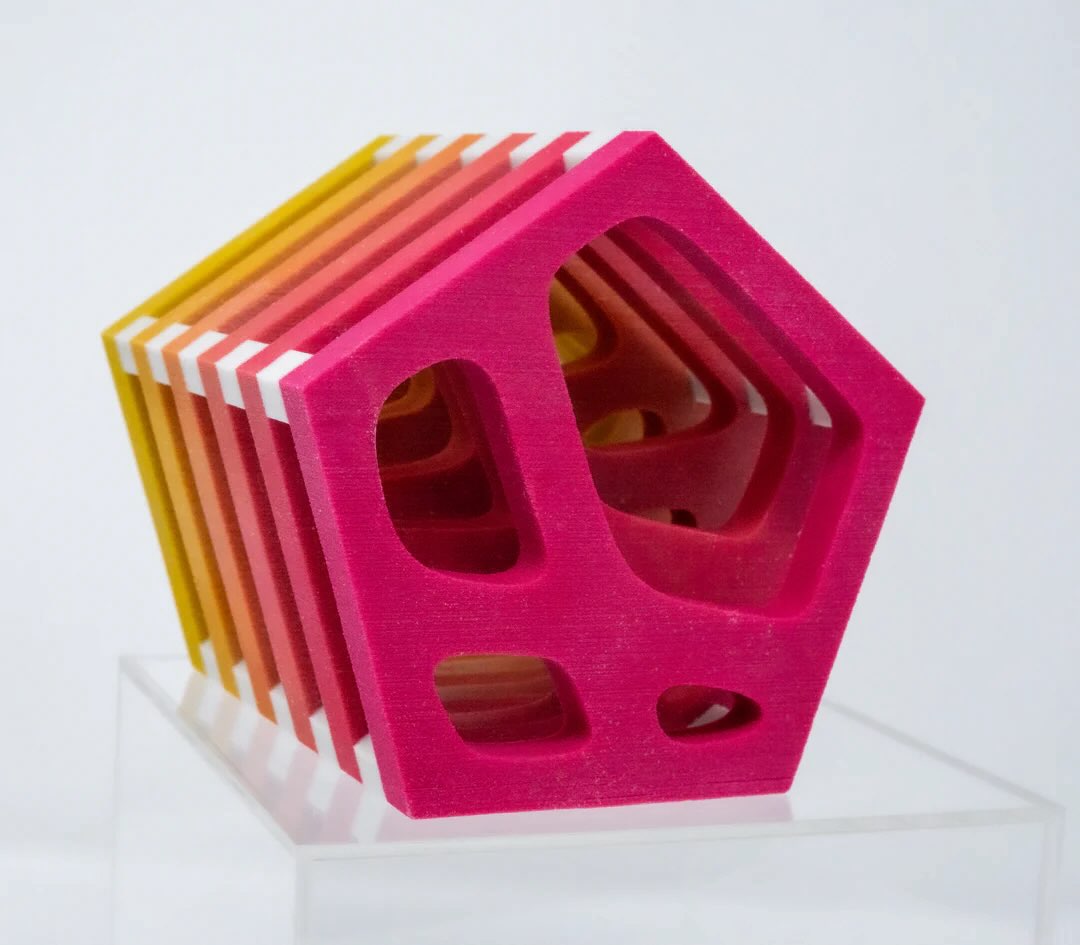}\vspace{-.05in}\\{\small photo by Valentina Ferro}\end{tabular}\vspace{.05in} 
 	\caption{Falling Inward by Colin Liotta, Mini Dahlias by Shawn Kemp, \\ and Pentagonal Window Evolution by Christopher Hanusa.}
 	\label{fig:inspiration}
\end{figure}

Three artworks stand out as inspiration for {\em Fractal Emergence}. (Figure~\ref{fig:inspiration}.) The first is \textit{Falling~Inward} by Colin Liotta \cite{Liotta}, a layered laser-cut wood visualization of the different rates of divergence of the Mandelbrot set. The second is Shawn Kemp's generative artwork inspired by Dahlia flowers, {\em Mini Dahlias} \cite{Kemp}, rendered physically using layered laser-cut card stock. The third is Hanusa's {\em Window Evolution} series \cite{Hanusa1}, a collection of 3D printed models with layers that show the evolution of a geometric scene over time. 

The {\em infinite trivalent tree} $\mathcal{T}$ is the tree with infinitely many vertices, all of degree three. This tree has been studied as the Cayley graph of the rank three free Coxeter group \cite{Goldman, LeeXu2019}. Goldman et al.\ \cite{Goldman} present a 3-edge-colored embedding of $\mathcal{T}$ in the Poincar\'e disk model of the hyperbolic plane~$\mathbb{H}^2$. {\em Fractal Emergence} is based on a different, symmetric and self-similar embedding of $\mathcal{T}$, shown in Figure~\ref{fig:fractalemergence}(b). In this paper, we share how following an iterative development process was crucial to achieving the final result.  

\vspace{.1in}
\section*{Iterative Development for {\em Fractal Emergence}}

Although it got its start in software engineering \cite{Larman}, {\em iterative development} has become a mainstay in many domains including hardware engineering, art, and design \cite{Asana}. The process involves cycling through ``design, build, and test'' loops, with each iteration refining some aspect of the product. The scale and scope of what is built and tested depends highly on what is being produced, as well as the resources available to the design team. Hardware development is expensive, so it is typically done at smaller scales.

In a professional environment, designing a hardware product can involve hundreds of people, ranging from designers and advertisers to engineers and technical experts. Prototypes can be produced in quantities rivaling small-scale production and testing can include  quantifying hundreds, if not thousands, of metrics.  The creation of {\em Fractal Emergence} replicated a significantly reduced version of this environment: each iteration involved building a single physical prototype, identifying opportunities for improvement using our judgment, and updating the design accordingly.

\begin{figure}
	\centering
	~~\includegraphics[width=6.25in]{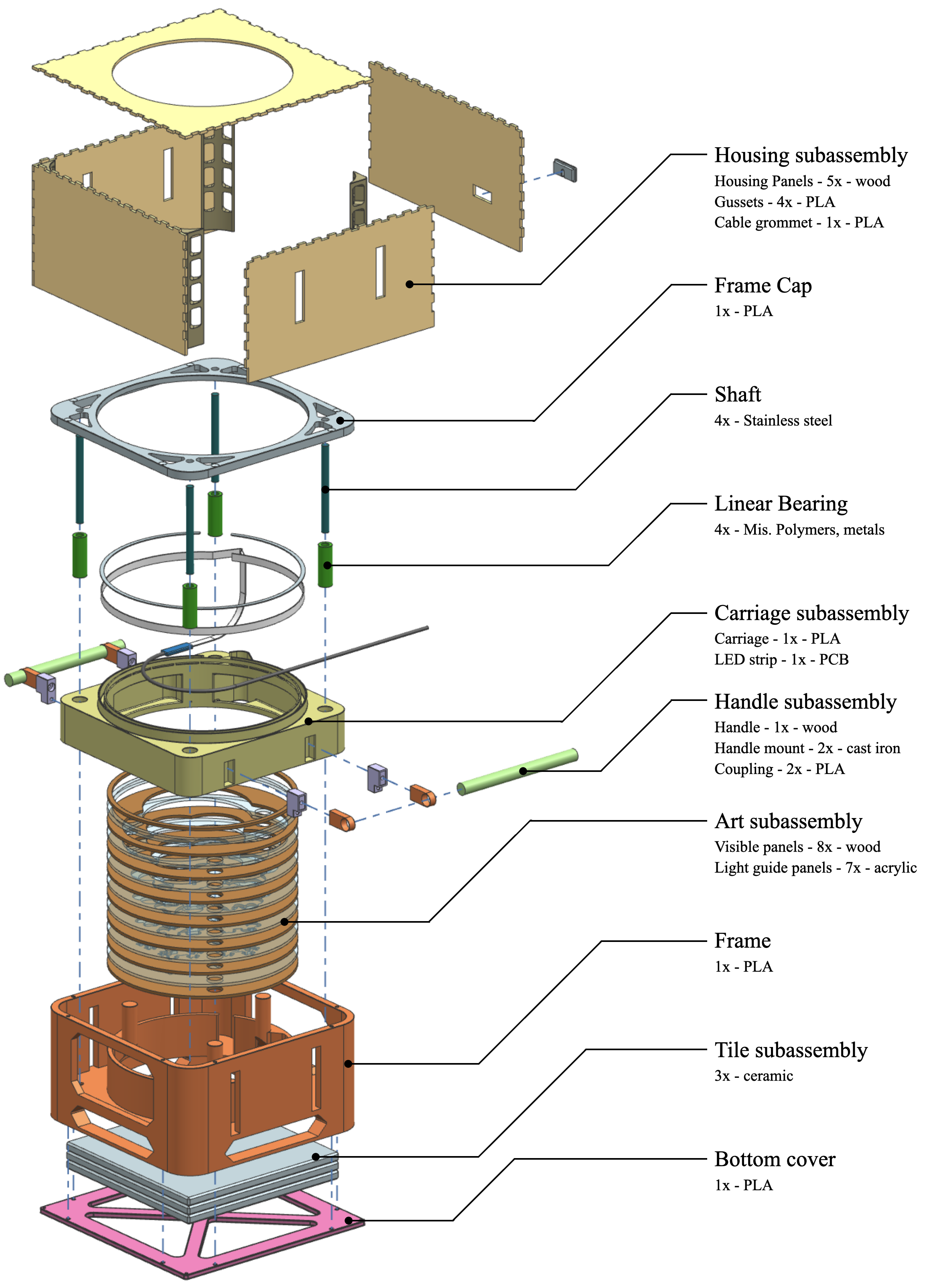}
	\caption{Exploded View for Fractal Emergence.}
	\label{fig:exploded}
\end{figure}

We used multiple software packages to design {\em Fractal Emergence}. {\em Mathematica} was used to model the fractal geometry because of its strength in algorithmic design and the ability to generate high quality visuals. From there, files were transferred to the 3D CAD software {\em NX}, where the bulk of the modeling took place. We generated a solid model and exported files for manufacturing.  Although not visible to the end user, there are many design features hidden inside {\em Fractal Emergence}. See Figure~\ref{fig:exploded} for details. 

We produced prototypes using rapid prototyping technologies such as 3D printing and laser cutting. These techniques are attractive because they enable the relatively inexpensive and fast production of parts when compared to other techniques such as CNC or injection molding. These qualities were crucial to our success because it enabled us to easily iterate the design, meaning we could explore what worked and what did not with only minor sunk costs. As with all technologies there are trade-offs involved; while low cost parts can be produced quickly, there are limits on the kinds of materials that can be used and the geometries of parts that can be produced. 

Our process for testing {\em Fractal Emergence} was atypical in the sense that all data we gathered was purely subjective. When evaluating a prototype we tested it against our guiding principles: the piece should be engaging, aesthetically pleasing, and well-crafted. This was accomplished by simulating the user experience of someone seeing the piece for the first time: we put ourselves in their shoes and tried to anticipate what their reaction would be by interacting with the prototype.

\begin{figure}[t]
\centering

    \includegraphics[height=0.85in]{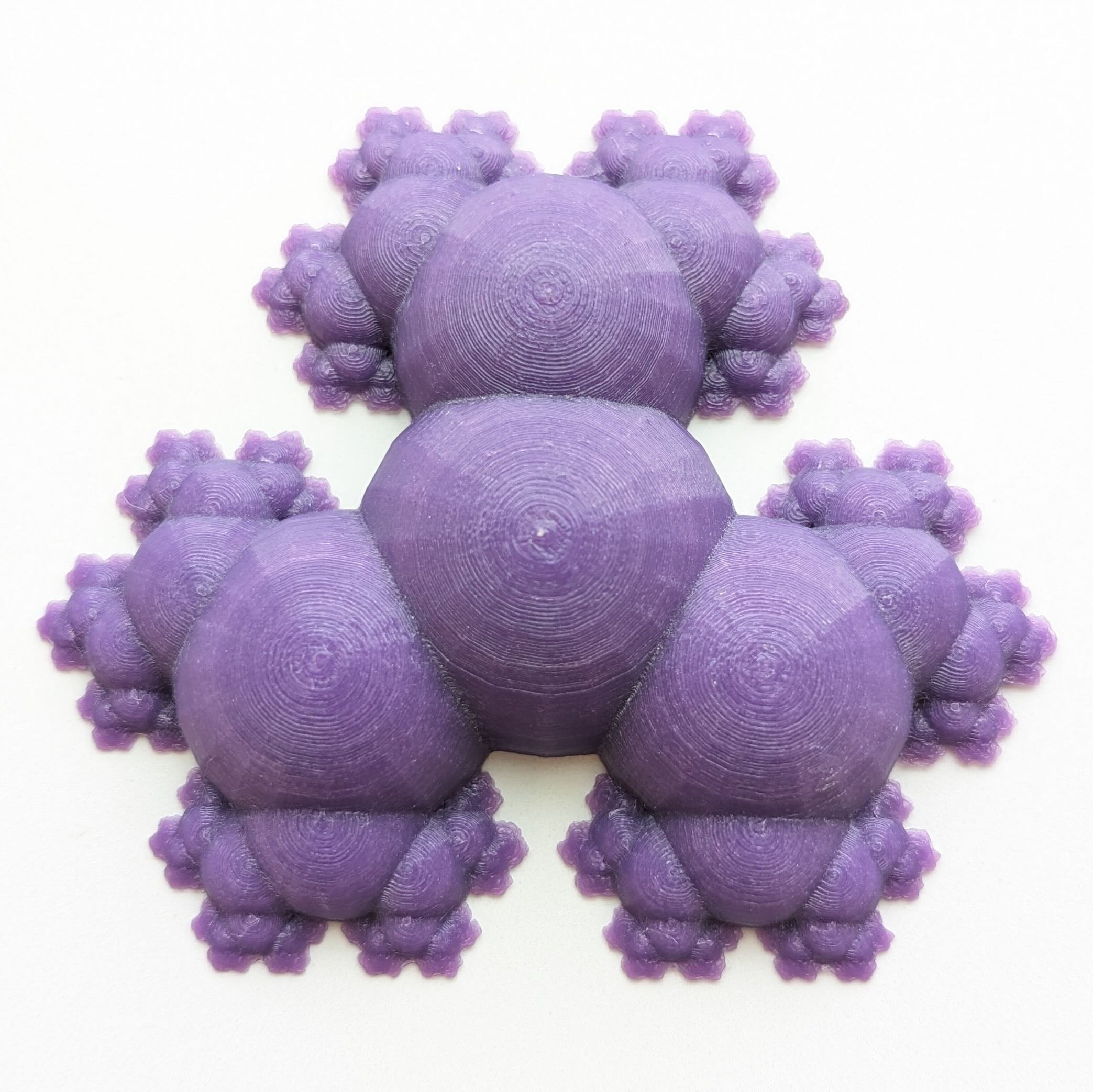} 
    \hspace{-0.95in}
    \raisebox{0.68in}{\colorbox{white}{\fbox{\small 0}}}
    \hspace{0.57in}
	\includegraphics[height=0.85in]{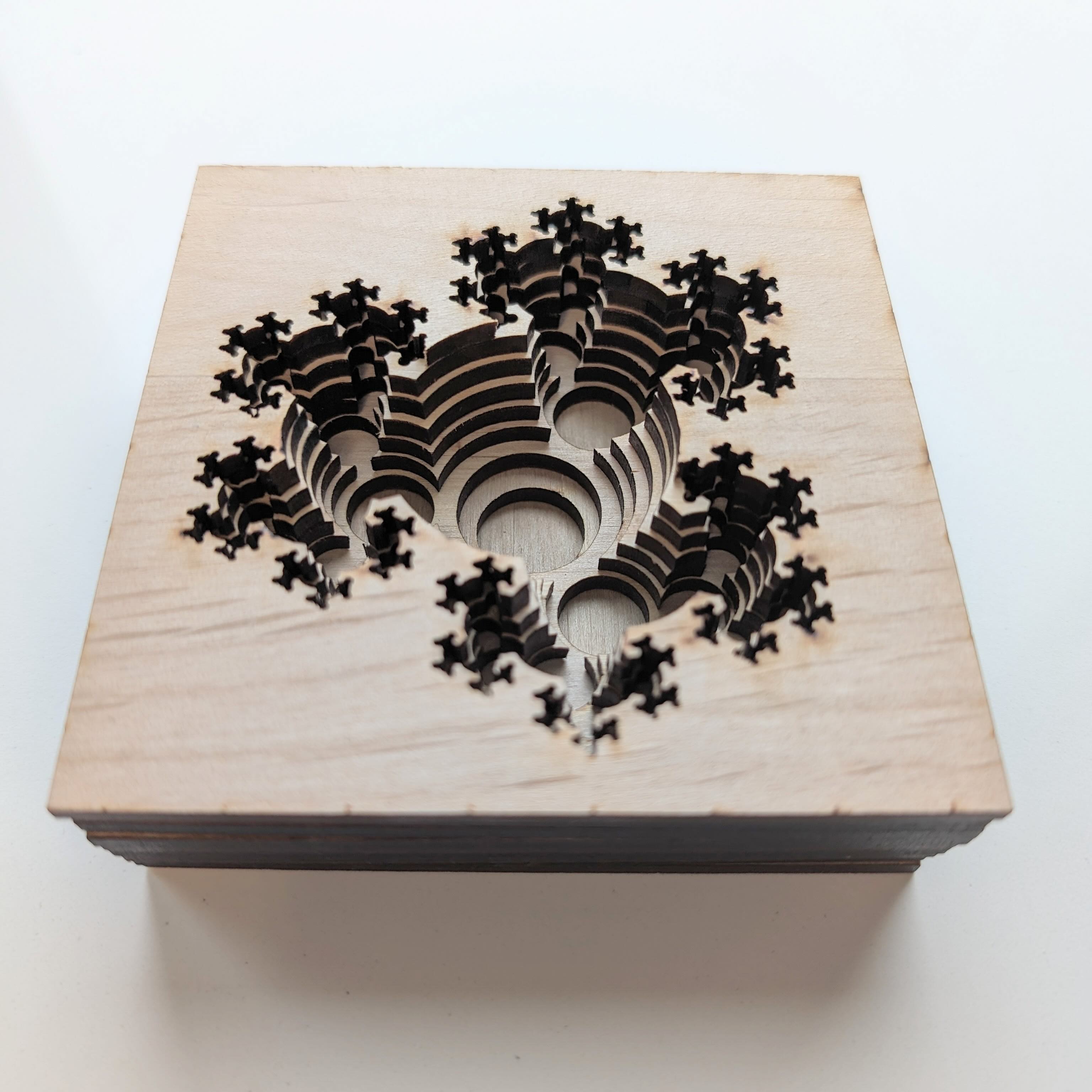}
    \hspace{-0.93in}
    \raisebox{0.68in}{\colorbox{white}{\fbox{\small 1.0}}}
    \hspace{4.92in}
    
\vspace{-0.8in}
    \includegraphics[height=0.85in]{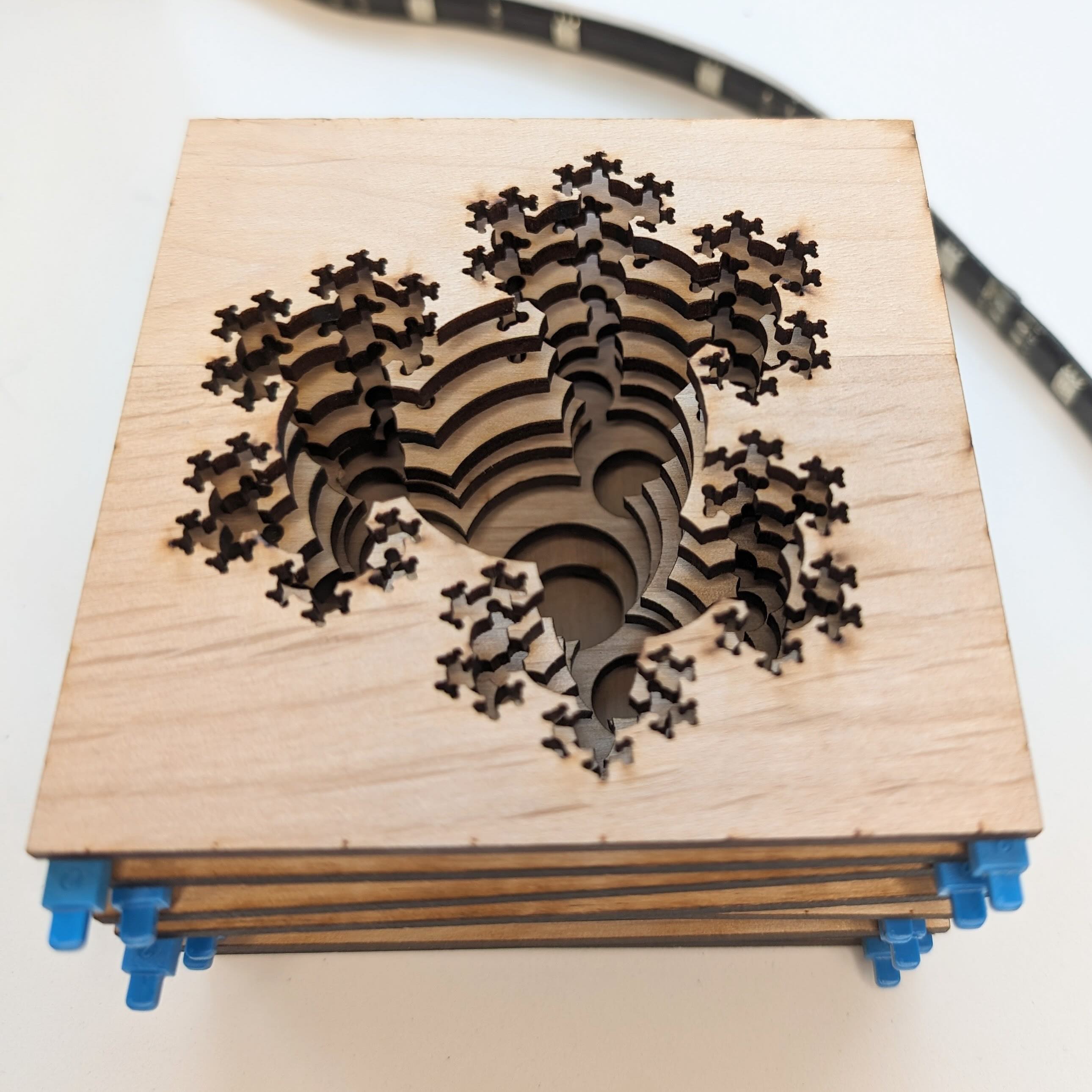}       
    \hspace{-0.95in}
    \raisebox{0.68in}{\colorbox{white}{\fbox{\small 1.1}}}
    \hspace{0.46in}
    \includegraphics[height=0.85in]{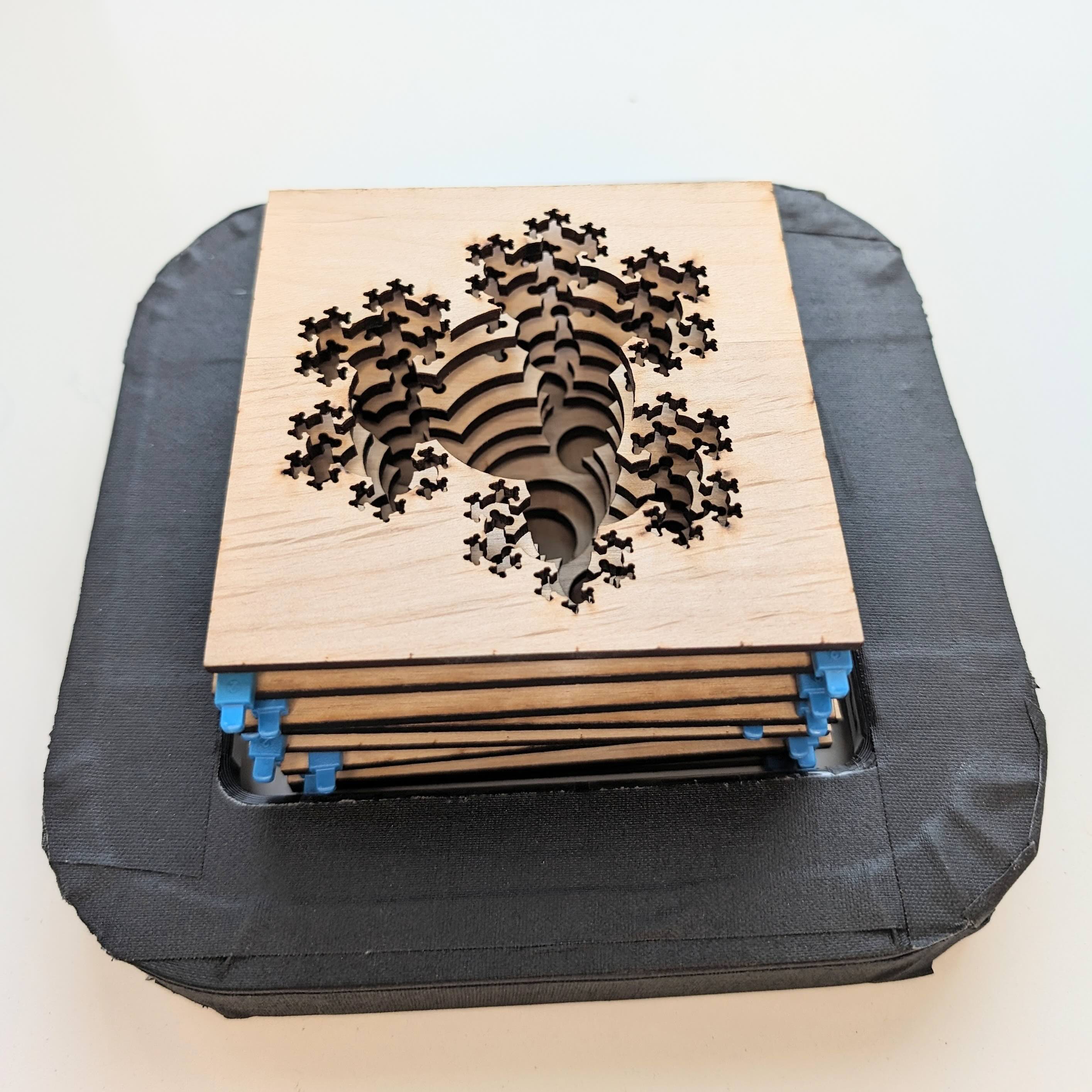}
    \hspace{-0.95in}
    \raisebox{0.68in}{\colorbox{white}{\fbox{\small 1.5}}}
    \hspace{0.48in}
    \raisebox{0.12in}{\includegraphics[height=1.45in]{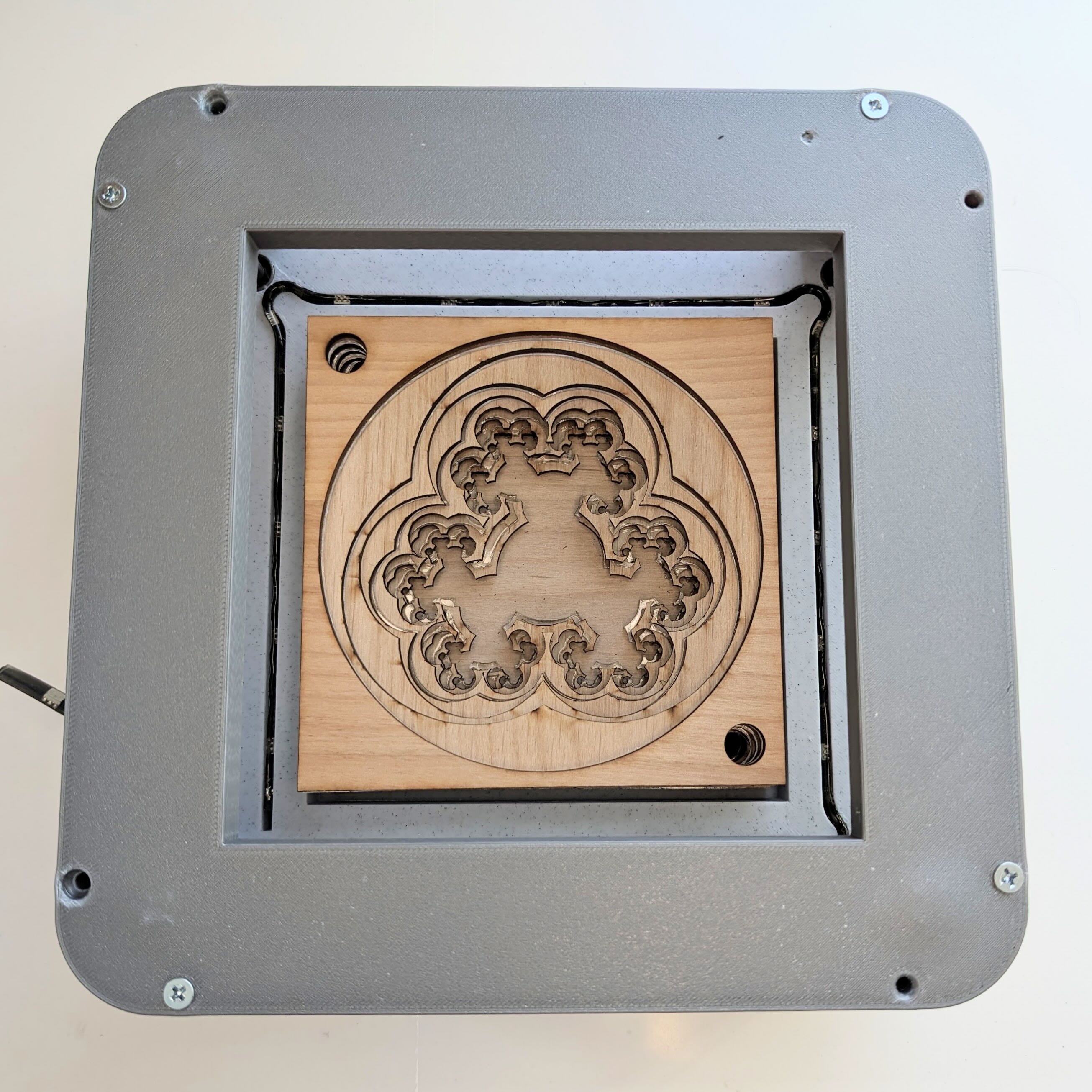}}
    \hspace{-1.54in}
    \raisebox{1.42in}{\colorbox{white}{\fbox{\small 2}}}
    \hspace{1.17in}
    \raisebox{0.12in}{\includegraphics[height=1.45in]{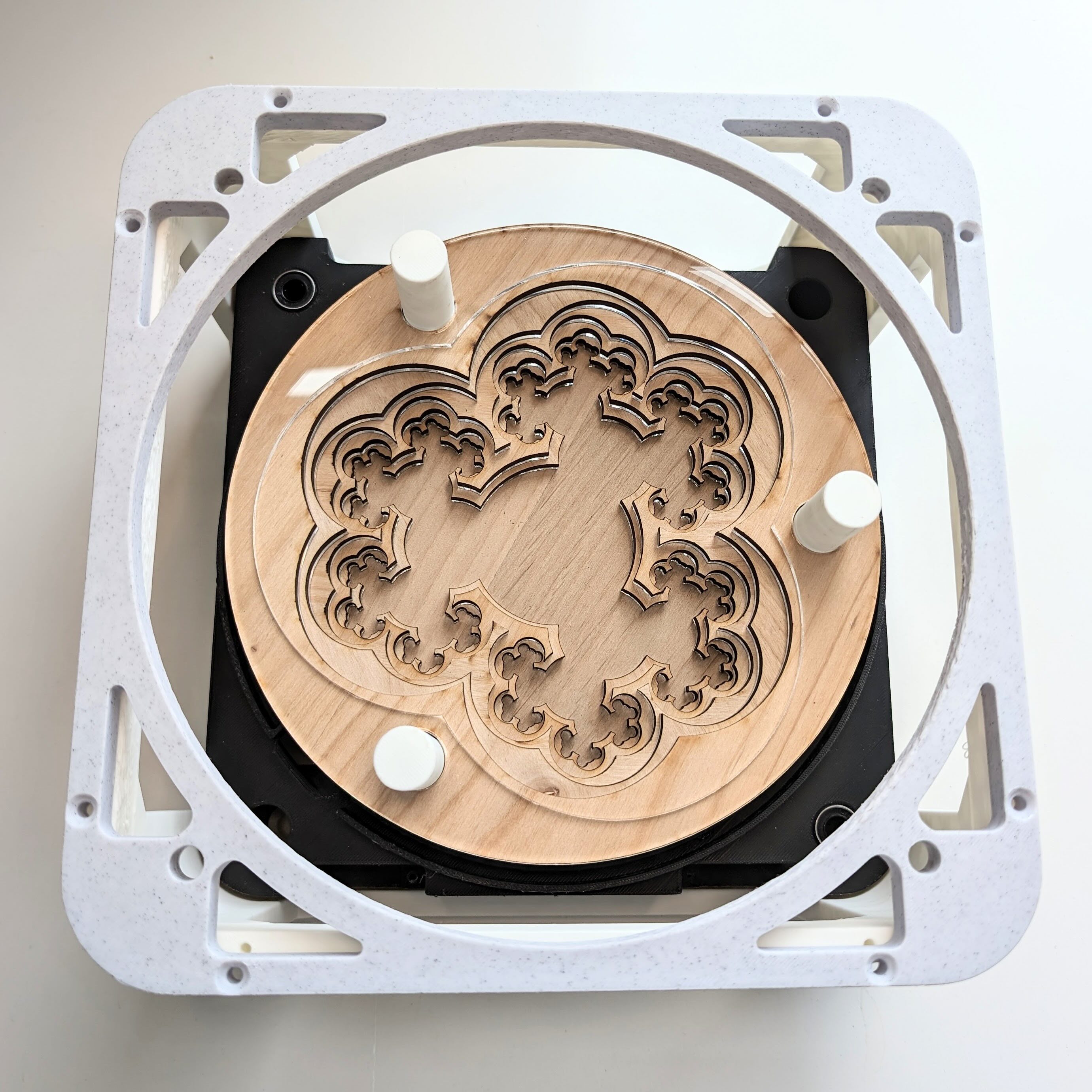}}
    \hspace{-1.54in}
    \raisebox{1.42in}{\colorbox{white}{\fbox{\small 3}}}
    \hspace{1.17in}
    \raisebox{0.12in}{\includegraphics[height=1.45in]{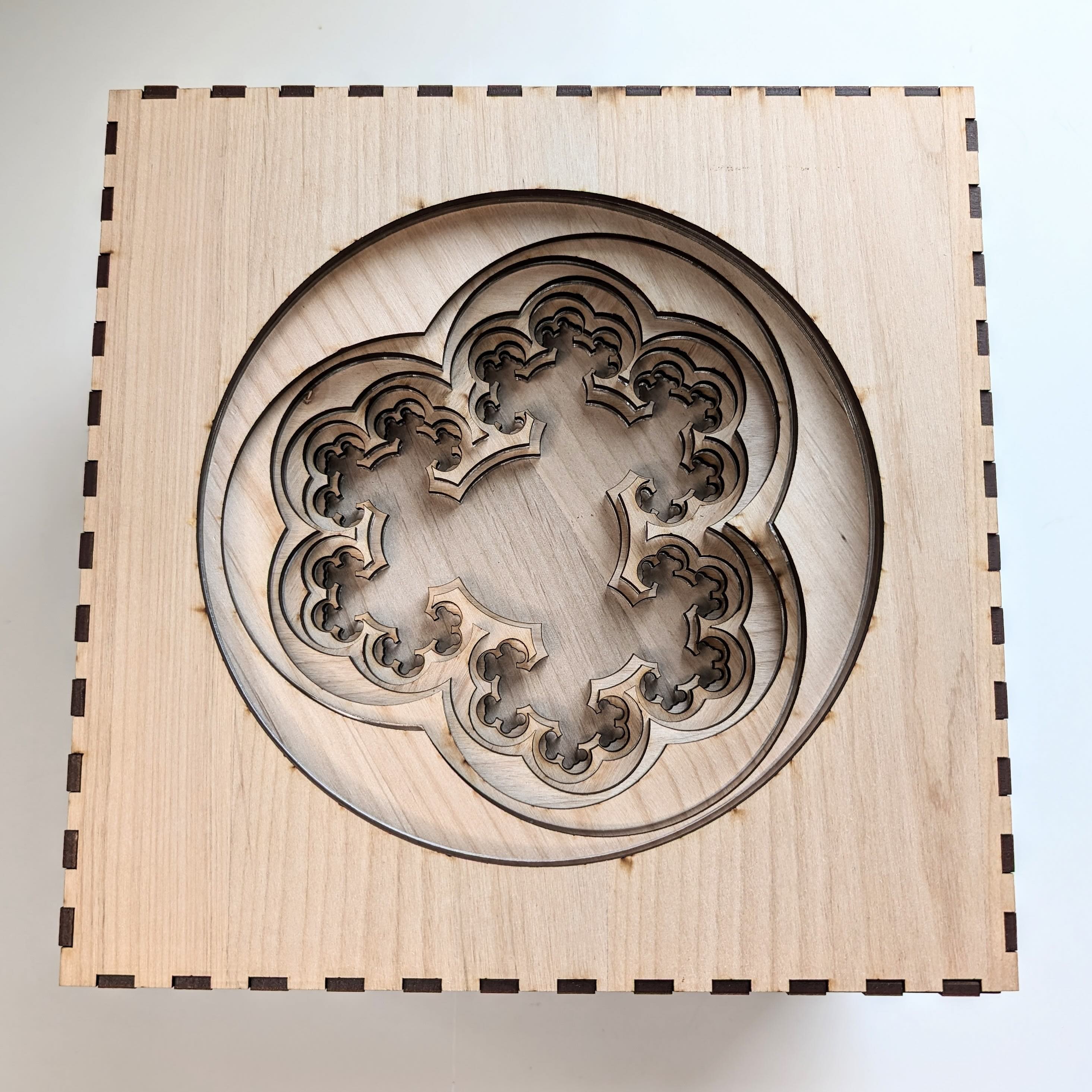}}
    \hspace{-1.54in}
    \raisebox{1.42in}{\colorbox{white}{\fbox{\small 4}}}
    \hspace{1.2in}
	\caption{The evolution of prototypes over time. Shown are versions 0, 1.0, 1.1, 1.5, 2, 3, and 4.}
	\label{fig:fractalovertime}
\end{figure}

\section*{Iterations of {\em Fractal Emergence}}

In this section we detail the steps of the iterative design process through which we arrived at our final piece. The reader is invited to follow along with the accompanying images in Figure~\ref{fig:fractalovertime}. 

\subsection*{Version 0: Initial Idea}

Version 0 of {\em Fractal Emergence} is the 3D print that Hanusa designed in 2017 that realizes the infinite trivalent tree $\mathcal{T}$ using overlapping spheres. We wondered if it was possible to capture its essence using laser-cut wood. To do so, we needed to formalize the structure of the embedding of $\mathcal{T}$ in the plane.

Our embedding starts by declaring one vertex~$v$ to be the root of $\mathcal{T}$ and subsequently choosing the locations of its three neighbors, the six vertices distance 2 away from $v$, the 12 vertices distance 3 away from $v$, and, in general, the $3\cdot 2^{d-1}$ vertices distance $d$ away from $v$. (See Figure~\ref{fig:fractalemergence}(b).) With the goal of having a symmetric and self-similar embedding, there are two parameters that can be adjusted: the ratio $p$ between the lengths of adjacent non-congruent edges (e.g.\ $vw$ and $wx$) and the angle $\theta$ between adjacent edges of the same length (e.g.\ $xy$ and $xz$). We chose $\theta=120^\circ$ to match the angle made by edges leaving $v$ and through trial and error arrived at $p\approx0.61$ to balance expanding the breadth of the embedding as much as possible and ensuring the embedding does not self-intersect. 

The tree is then thickened to a shape with positive area by positioning a disk at each vertex with a radius that depends on its distance $d$ away from $v$. Requiring self-similarity restricts the choices to two parameters: the radius $r$ of the disk at $v$ and the ratio $q$ of the radii of disks centered at adjacent vertices (e.g.\ $w$ and $x$). These values were honed during the iterative design process.

\medskip
\subsection*{Version 1: First Steps with Prototyping}

\begin{figure}[t]
	\centering
\raisebox{1.0in}{(a)}~\includegraphics[height=1.15in]{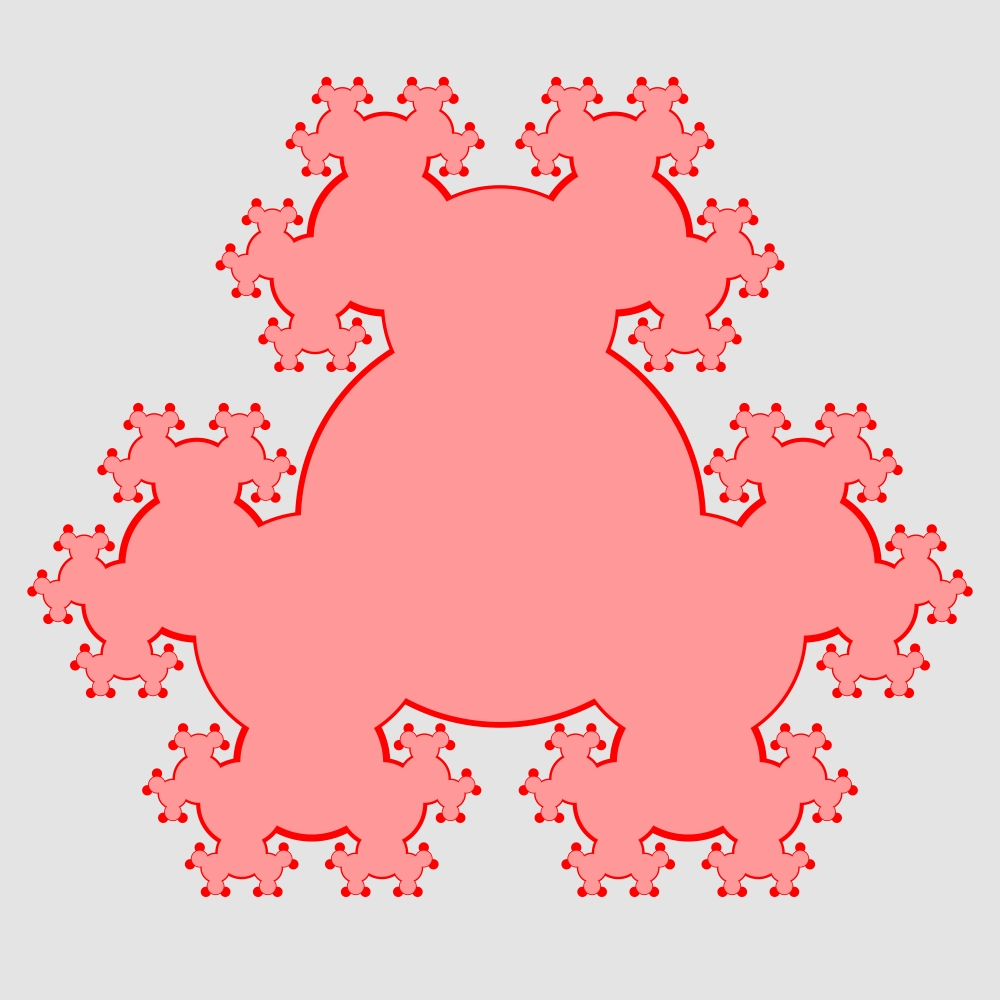}
~~\raisebox{1.0in}{(b)}~\includegraphics[height=1.15in]{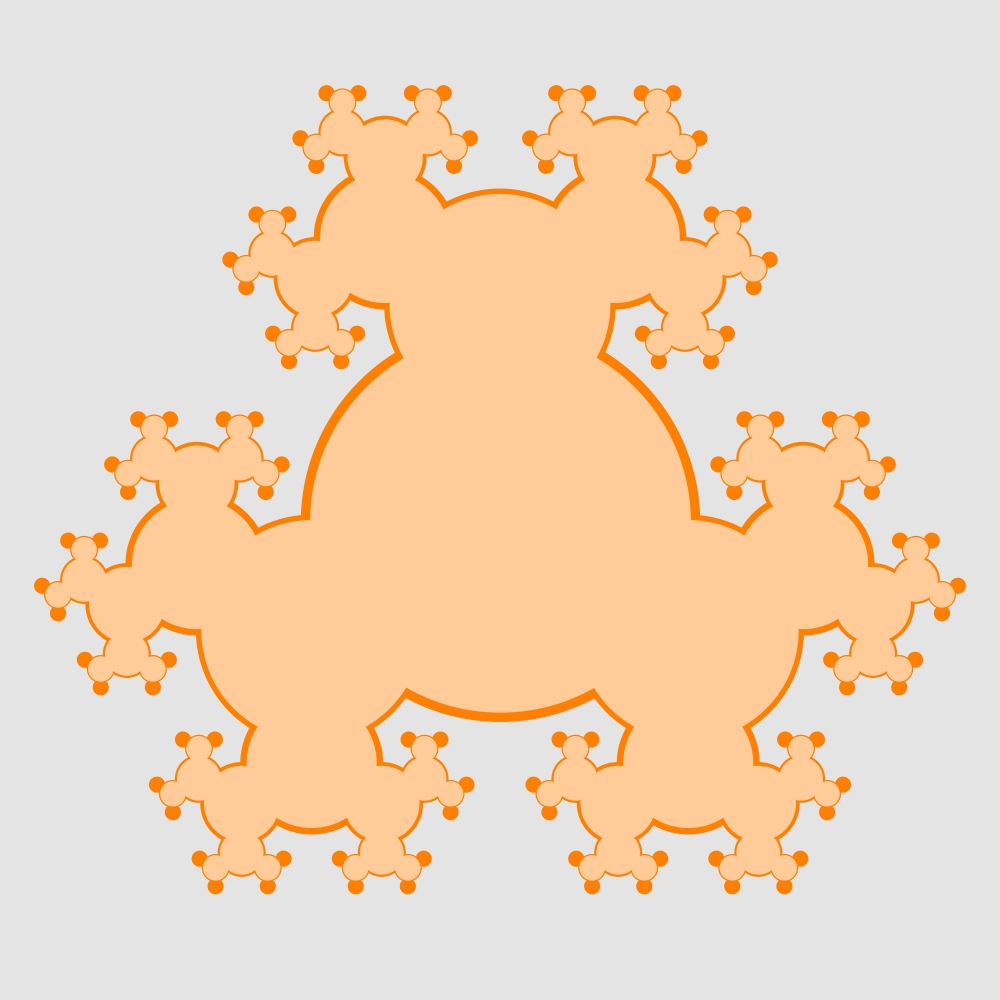}
~~\raisebox{1.0in}{(c)}~\includegraphics[height=1.15in]{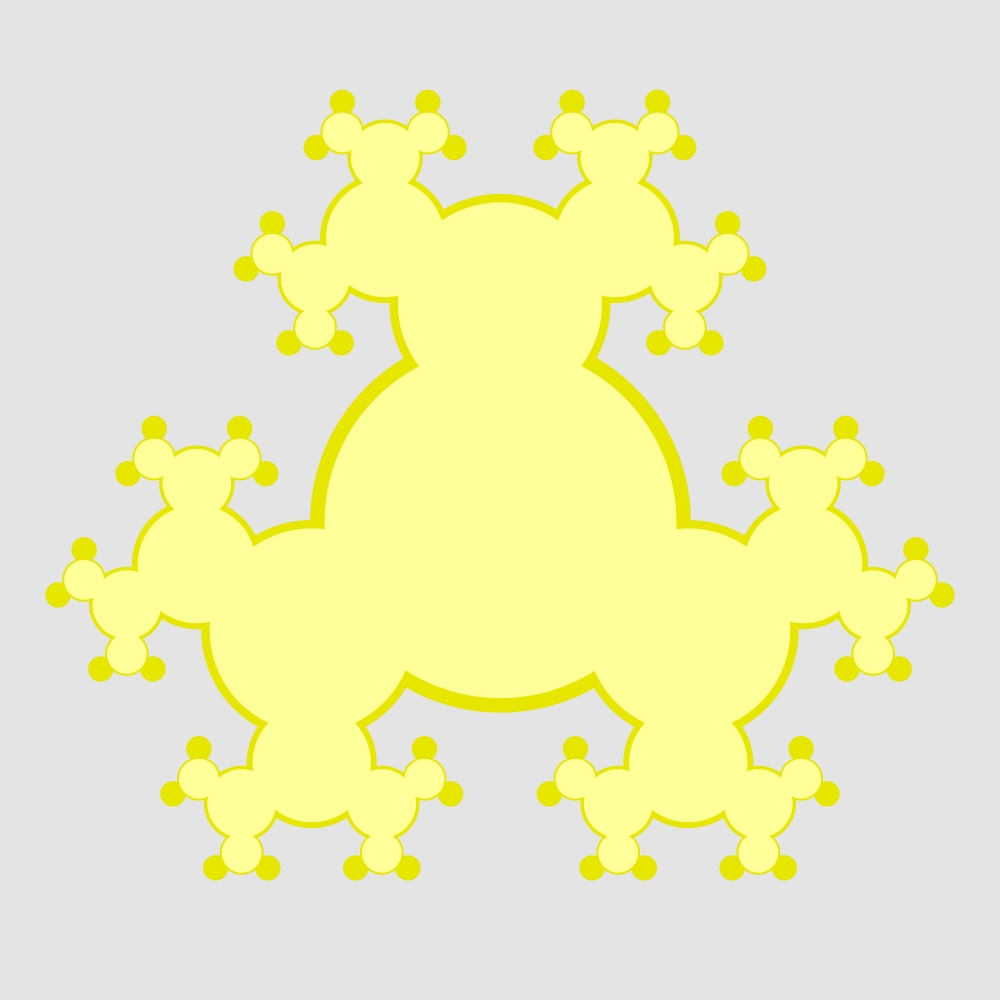}
~~\raisebox{1.0in}{(d)}~\includegraphics[height=1.15in]{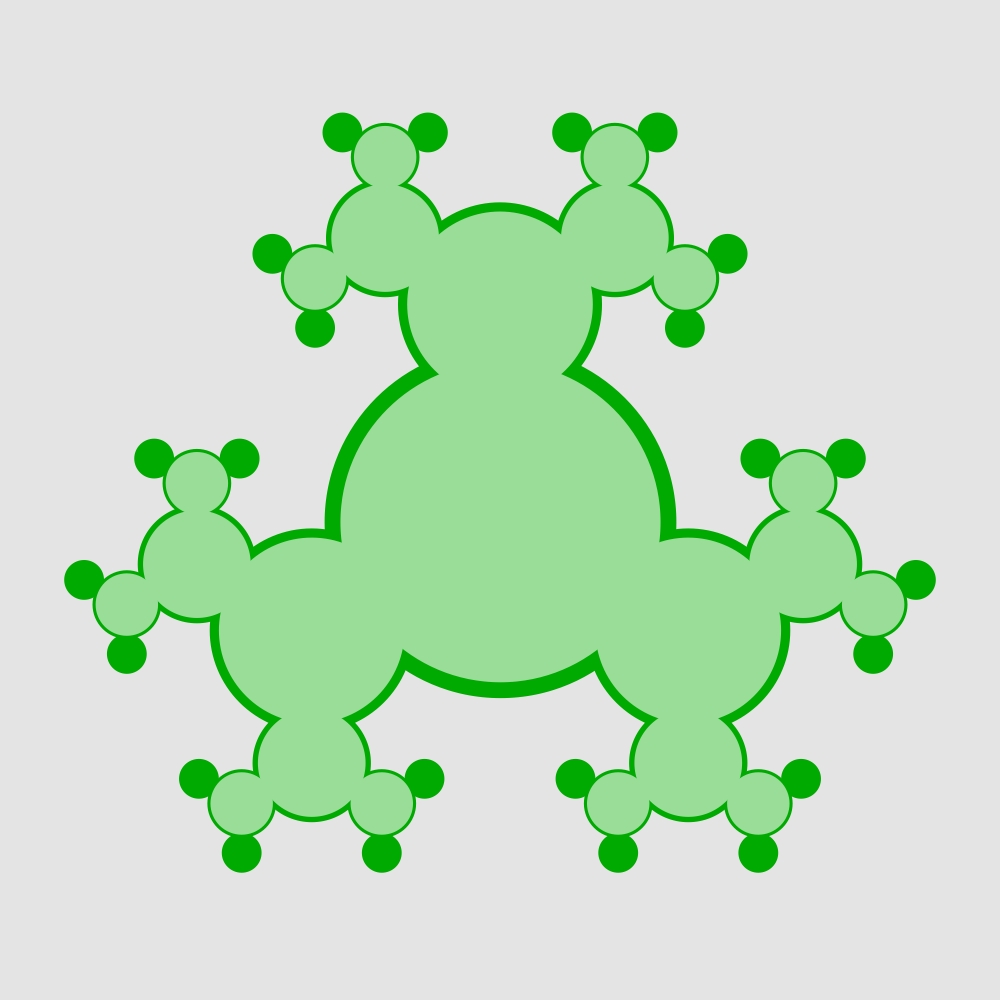}~~~~~\phantom{.}\medskip

\raisebox{1.0in}{(e)}~\includegraphics[height=1.15in]{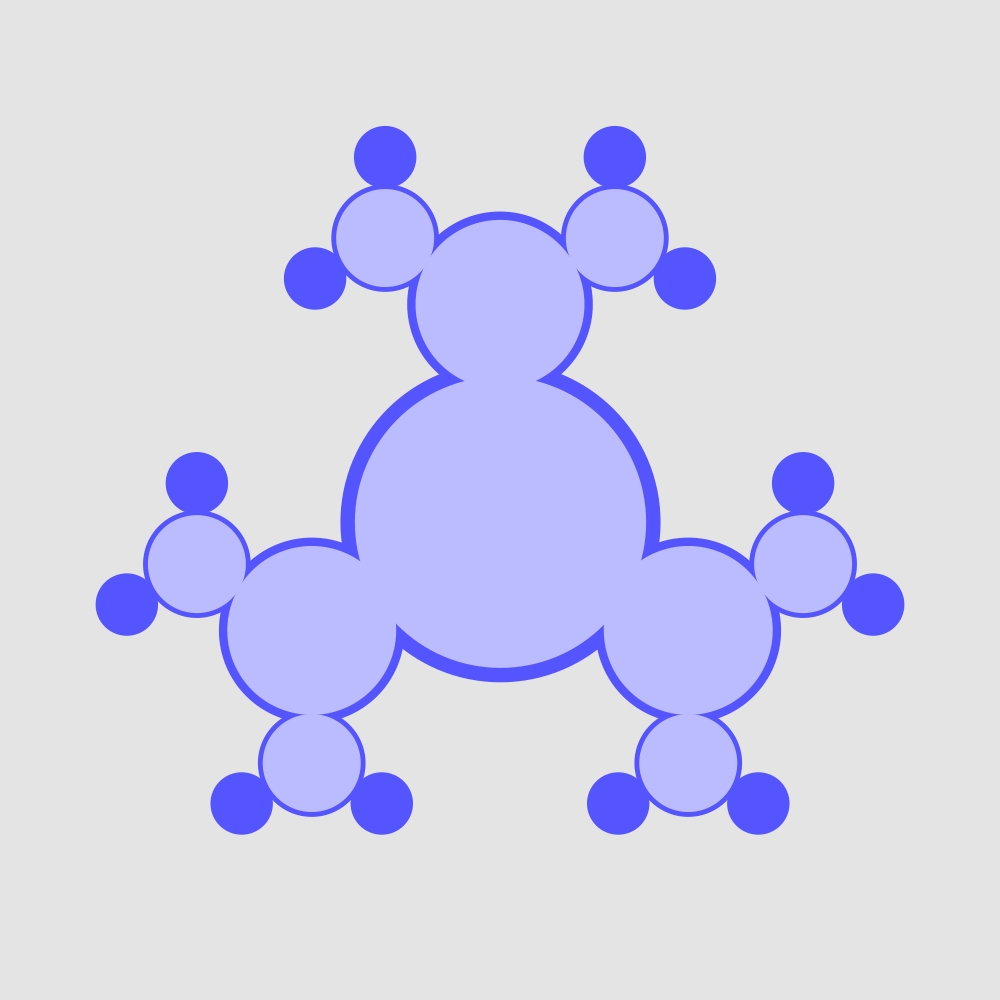}
~~\raisebox{1.0in}{(f)}~\includegraphics[height=1.15in]{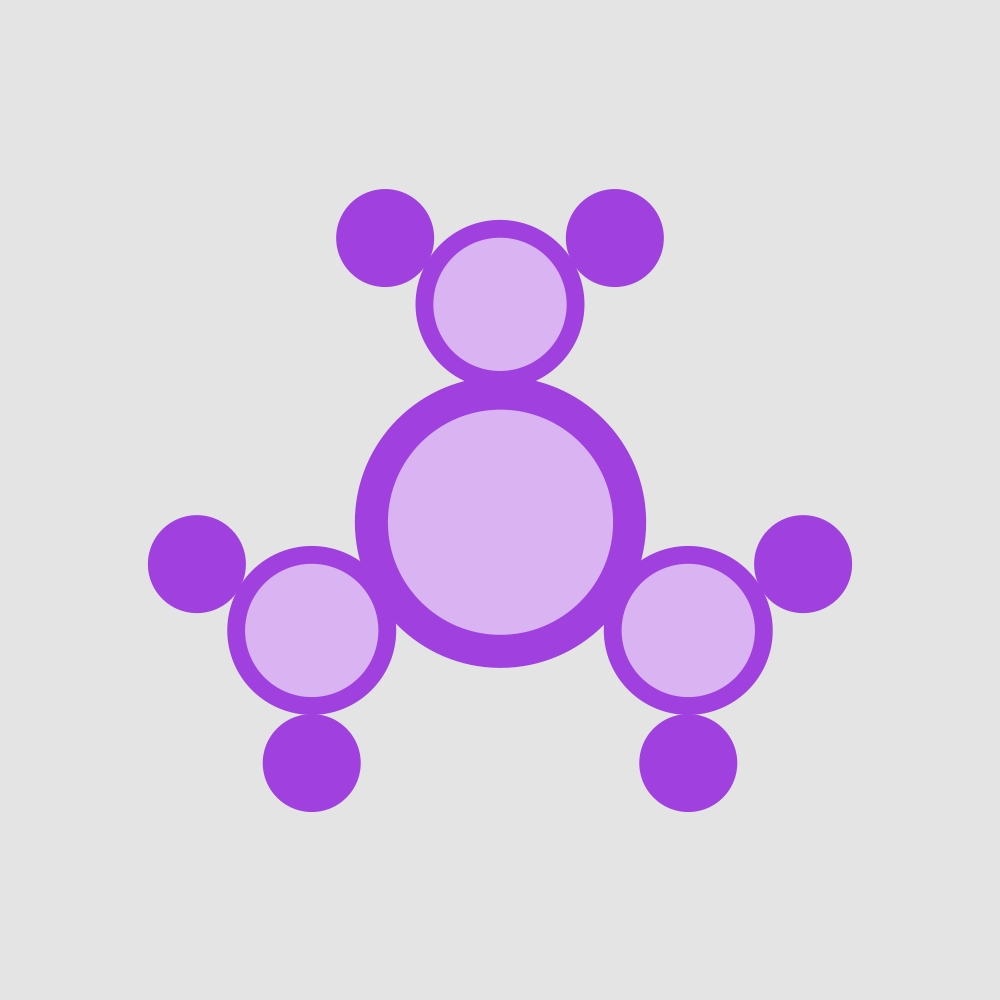}
~~\raisebox{1.0in}{(g)}~\includegraphics[height=1.15in]{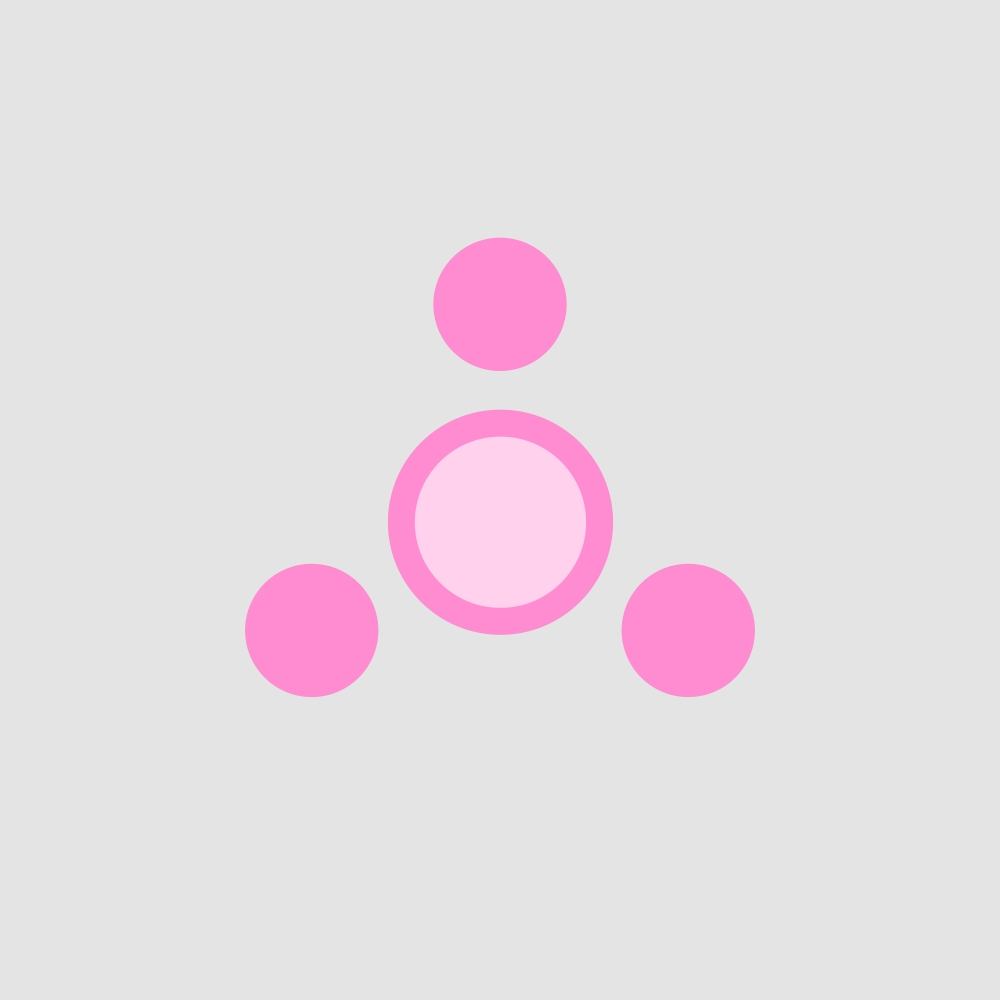}
~~\raisebox{1.0in}{(h)}~\includegraphics[height=1.15in]{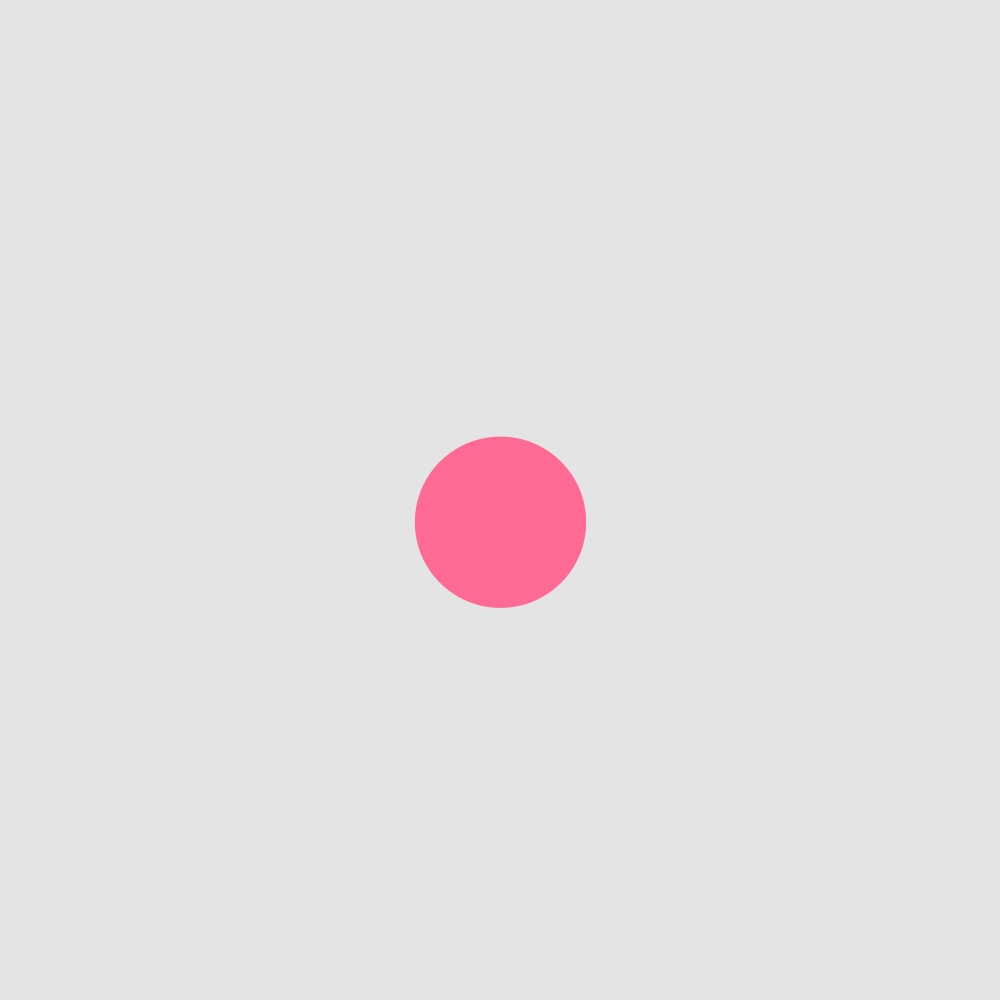}~~~~~\phantom{.}

	\caption{The layers of version 1 of Fractal Emergence from top to bottom. The colored region has been cut out of the layer; the darker part of the colored region is the visible part of the layer beneath.}
	\label{fig:layers-v1}
\end{figure}

In attempting to recreate the 3D model using laser-cut wood, we imagined each layer of the wood would represent a 2D cross section. However, slicing the model at equally spaced heights does not recover the fractal nature of the piece. Instead, we created one layer for each iteration of the fractal process by including larger and larger neighborhoods of the root vertex $v$. (See Figure~\ref{fig:layers-v1}.) We arranged the wood layers in a stack; this formed version 1.0 of our work. 

Sitting with the stack of laser-cut wood started a conversation, revolving around questions that emerged from our guiding principles. We realized that the interior of the artwork was particularly dark, making it difficult to see all of the intricate details and ultimately not engaging. Taking the cue from the {\em Window Evolution} series, we added spacing between the layers to let in some light. This worked! The layers became more visible. Suddenly, we realized that the visual would be enhanced if additional light was added through the gaps created by the spacers. We procured a store-bought LED strip and waved it along the outside of the prototype, forming version 1.1 of our work. At that point, we knew we had something engaging. 

As we moved the LED strip up and down, new, unanticipated questions emerged with the end user experience in mind, including ``How will the lights be powered in an integrated piece?''. This led to the design and 3D printing of a rigid square structure to house the LEDs and provide some consistency to how the lights move. This wood stack and LED structure formed version 1.5 of our work. 

We realized that our goal of engagement was more specifically a goal that the user be given the agency to experience and {\em interact} with the visualization for themselves. This clarified the principle that guided us through the rest of the process: {\em Make the user experience as reliable and powerful as possible.} 

Improving the reliability of the user experience would require changing the construction of the LED structure to ensure that its movement lights each horizontal layer one by one.  Moreover, the initial mathematical construction of the layers, in which disks corresponding to the same vertex in $\mathcal{T}$ are centered in the same position across all levels led to a situation where the amount of wood that is illuminated at each level is minimal. (See the darker part of the colored regions in Figure~\ref{fig:layers-v1}.) To make the user experience more powerful, the mathematical construction would need to be re-imagined to ensure that the amount of wood that is revealed at every layer is substantial. We would implement these changes in the next prototype.

\medskip
\subsection*{Version 2: A Major Step Forward}

Version 2 involved a square 3D printed cage, a square 3D printed carriage that moved up and down with linear ball bearings on metal rods, and an updated mathematical visualization. 

\begin{figure}
	\centering

\raisebox{1.in}{(a)}~\includegraphics[height=1.15in]{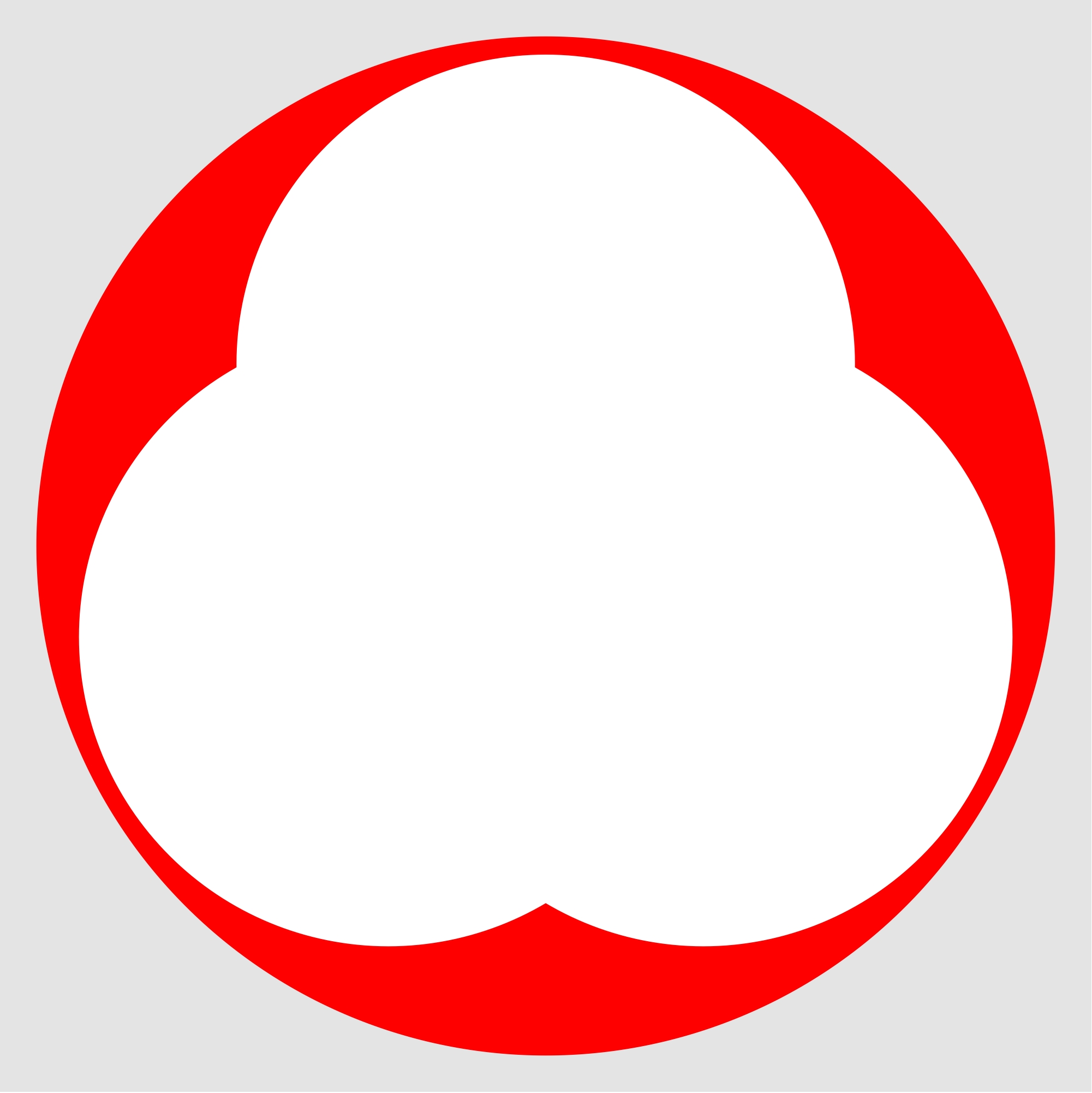}
~~\raisebox{1.in}{(b)}~\includegraphics[height=1.15in]{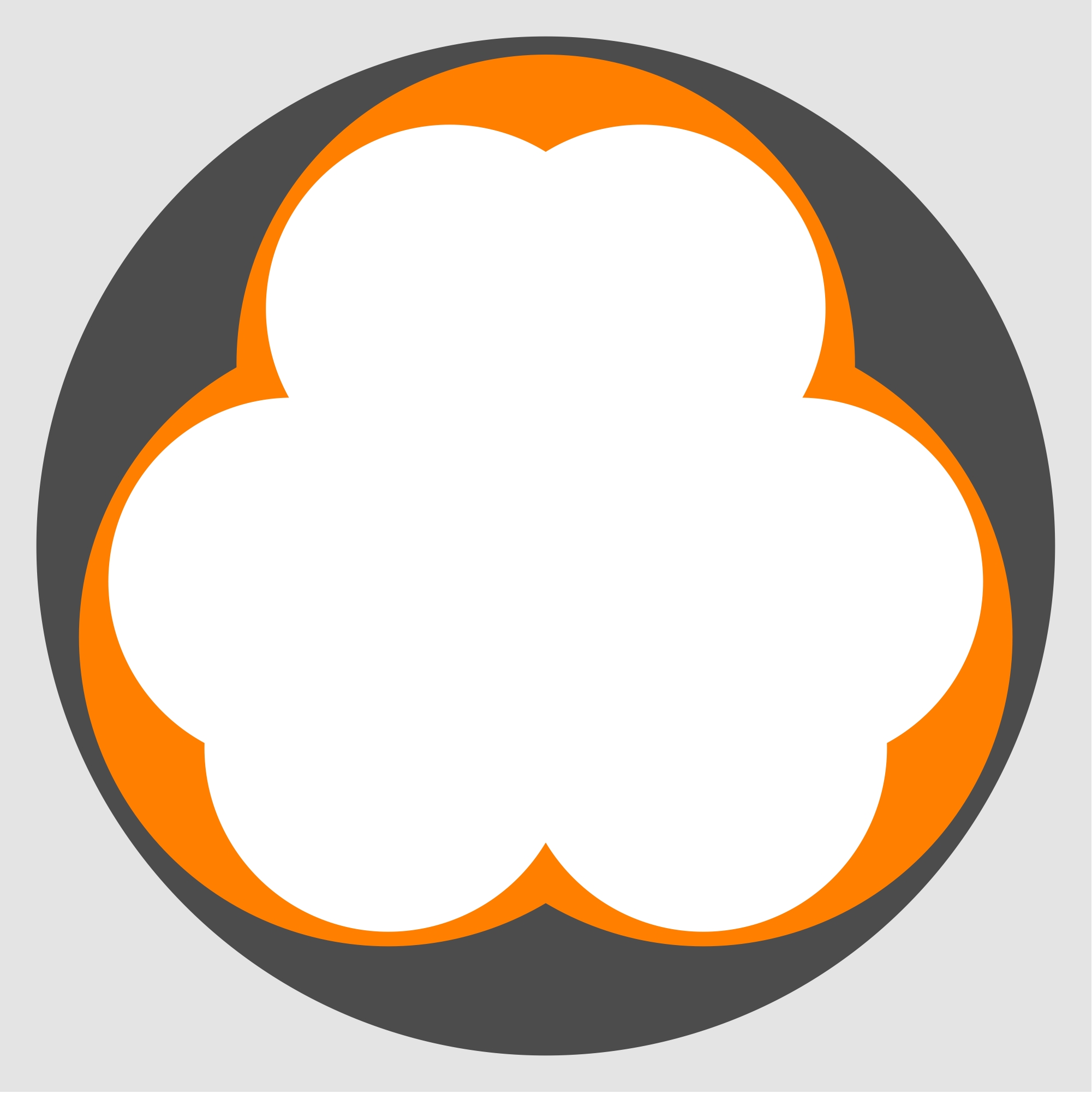}
~~\raisebox{1.in}{(c)}~\includegraphics[height=1.15in]{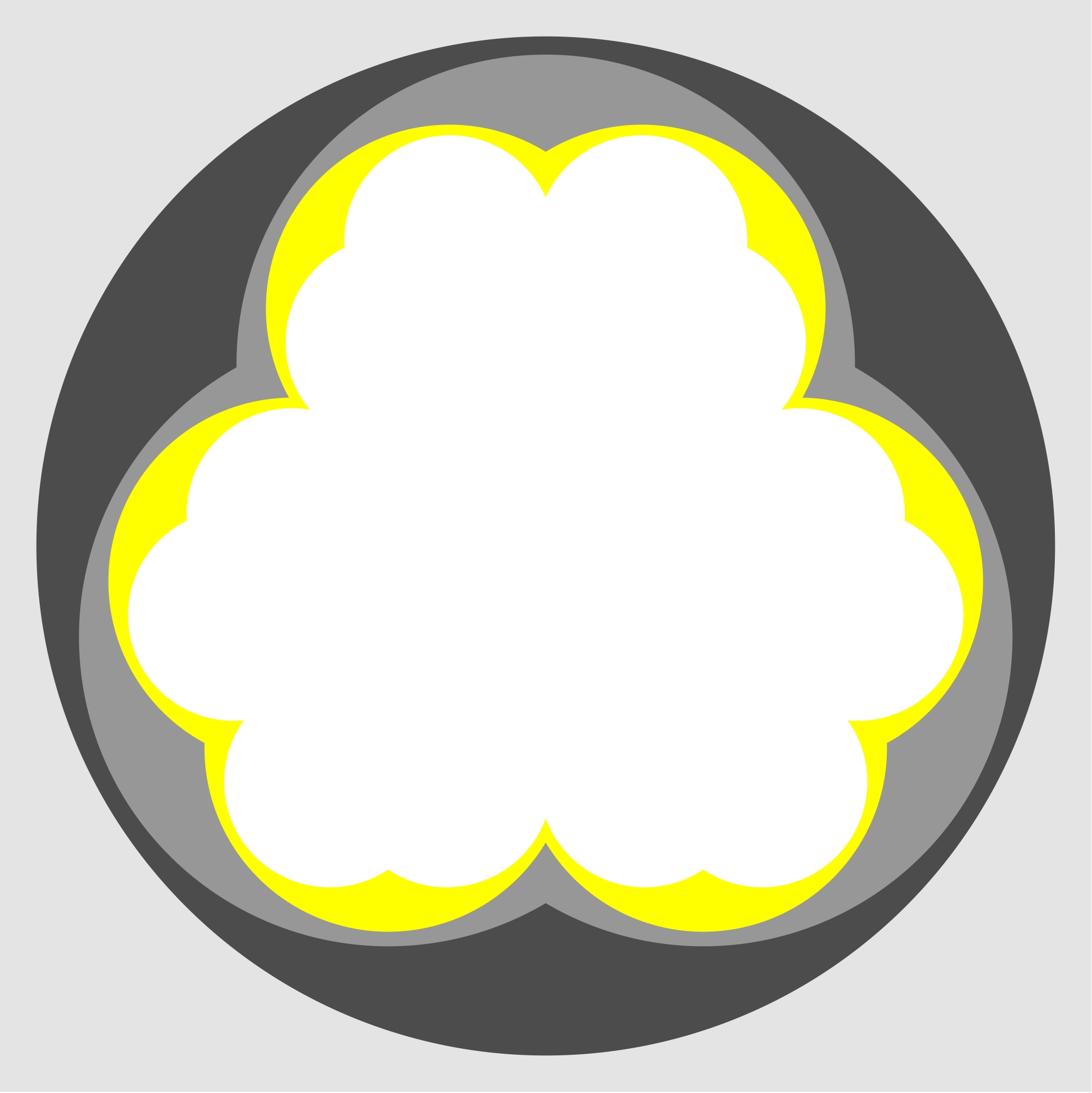}
~~\raisebox{1.in}{(d)}~\includegraphics[height=1.15in]{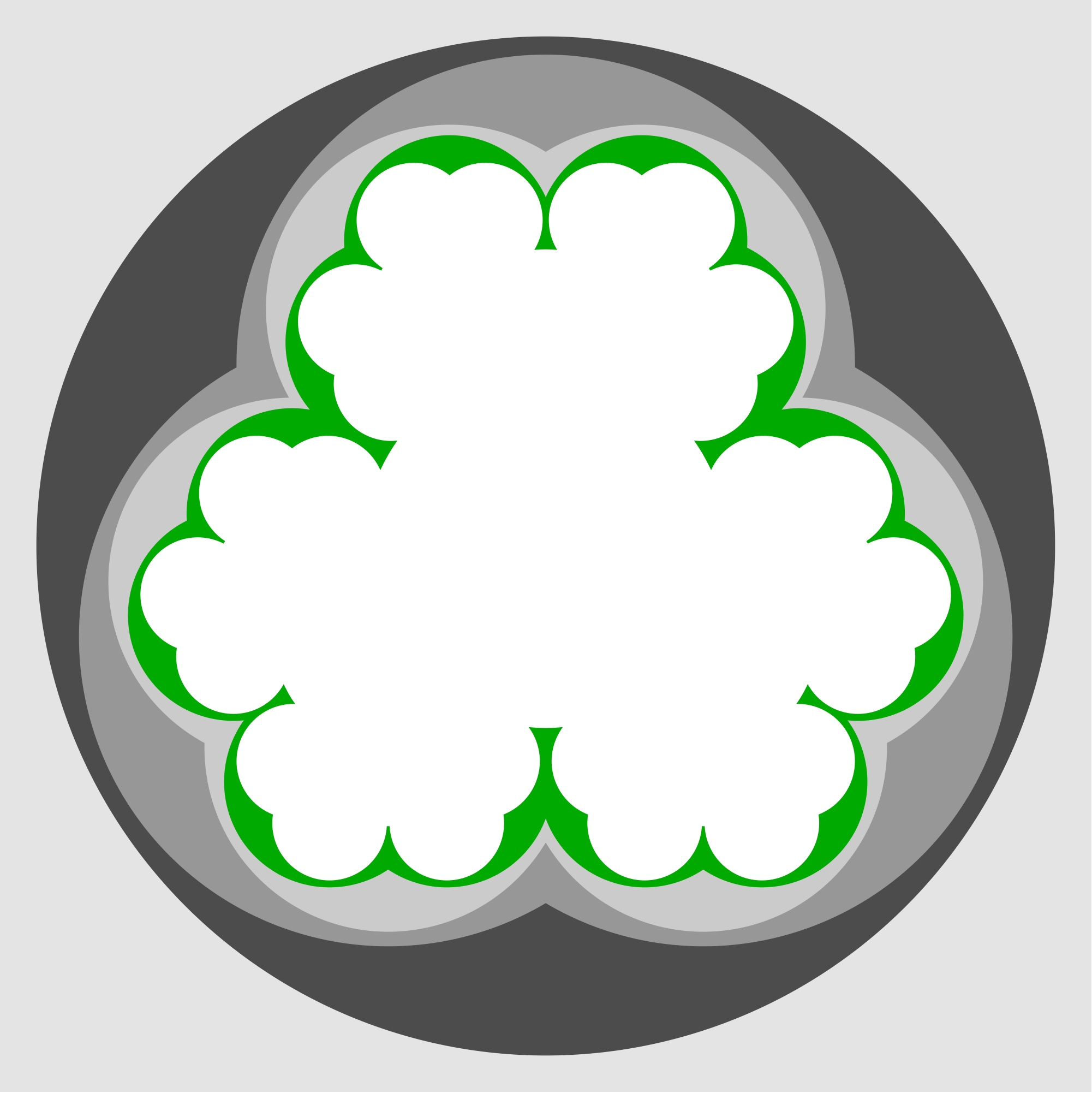}~~~~~\phantom{.}\medskip

\raisebox{1.in}{(e)}~\includegraphics[height=1.15in]{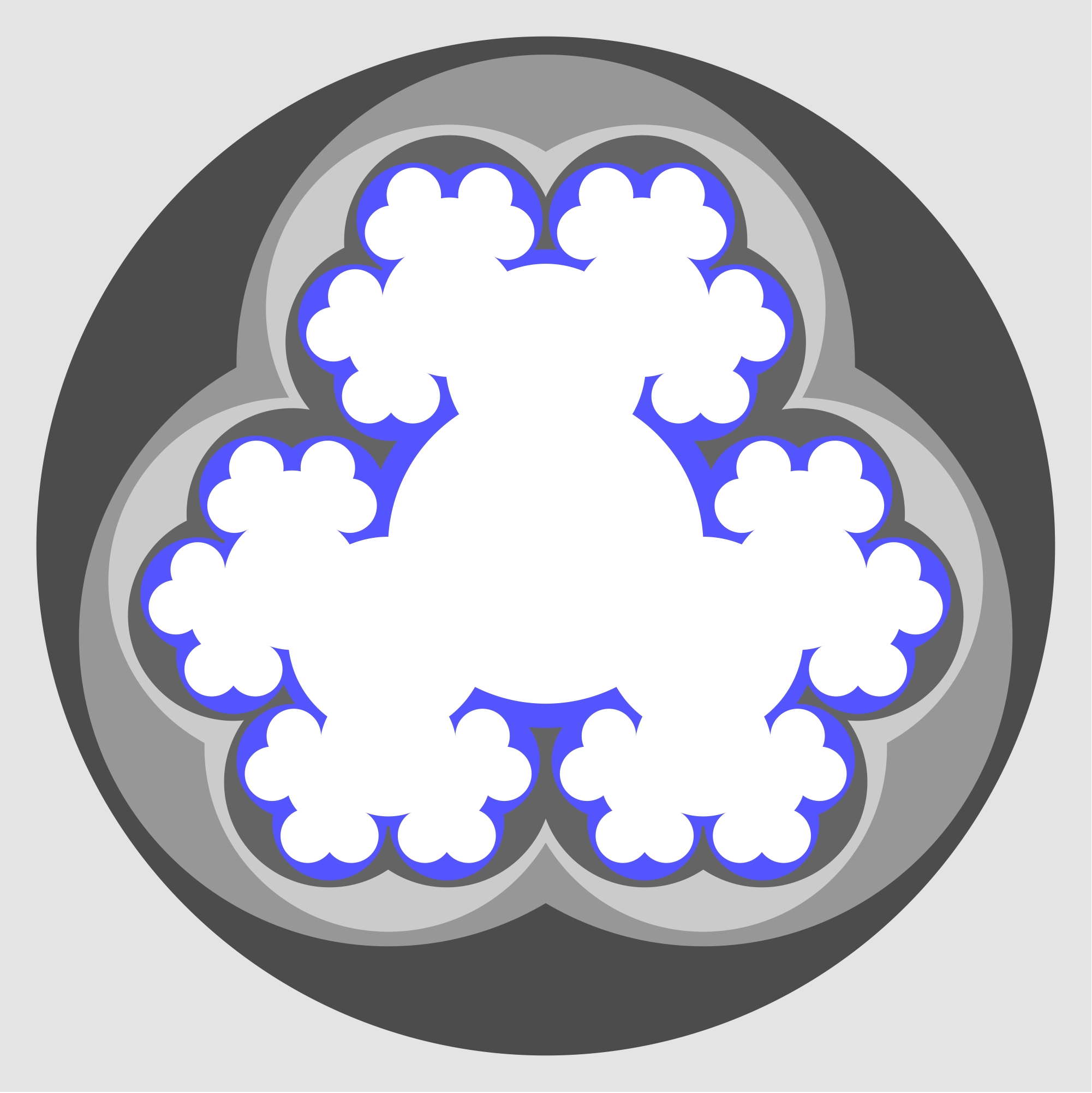}
~~\raisebox{1.in}{(f)}~\includegraphics[height=1.15in]{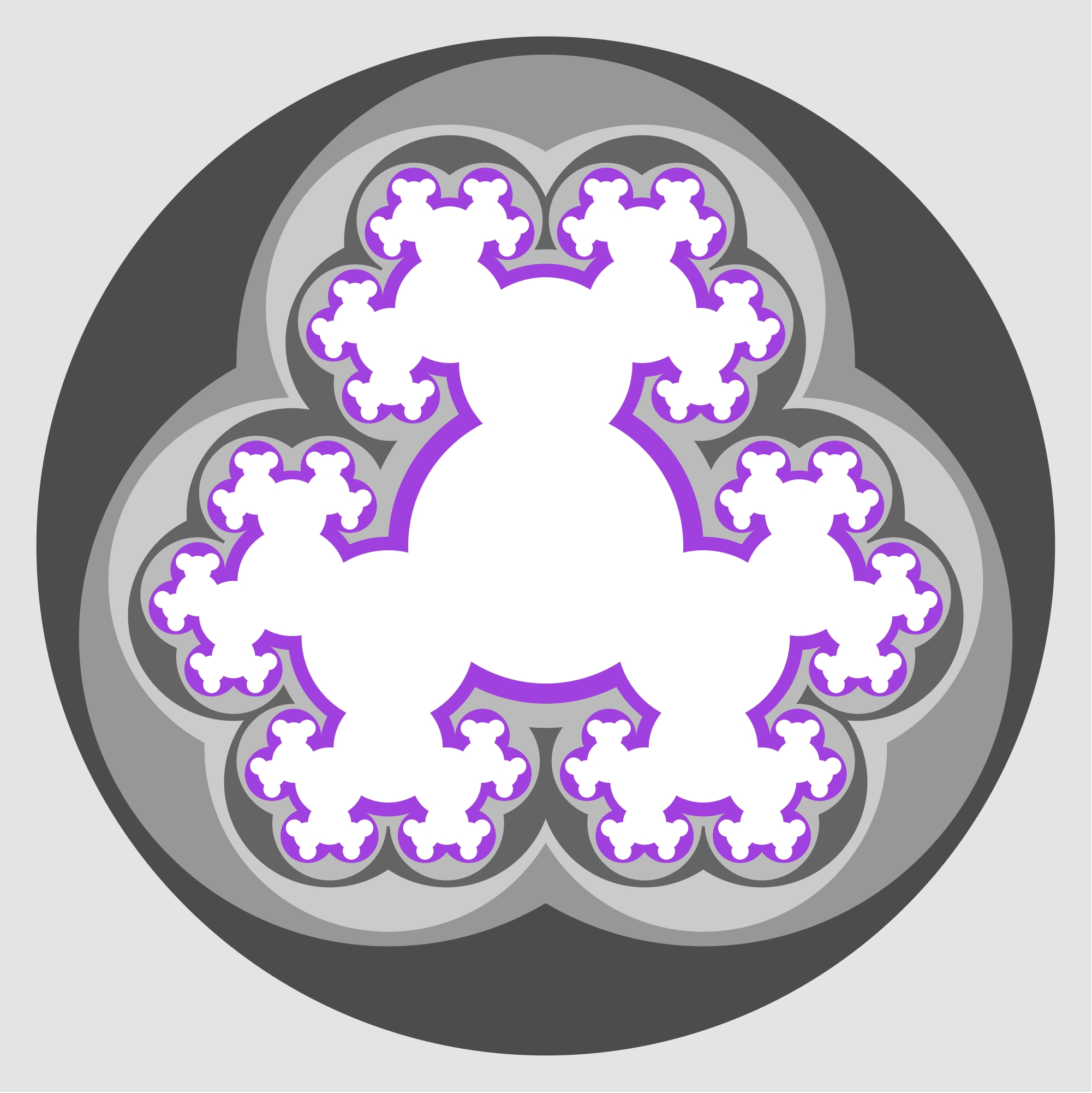}
~~\raisebox{1.in}{(g)}~\includegraphics[height=1.15in]{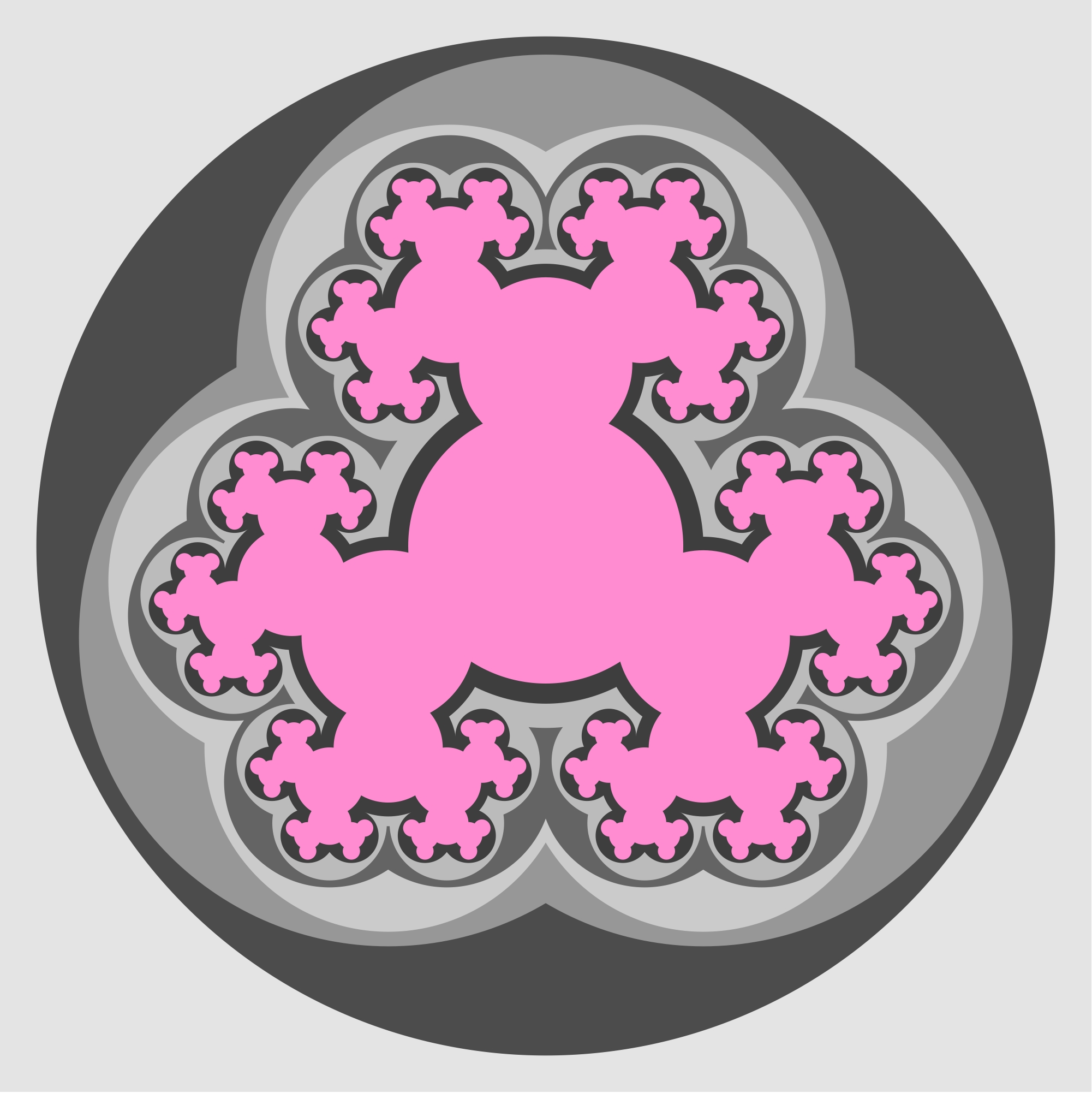}~~~~~\phantom{.}

	\caption{The layers of versions 2--5 of Fractal Emergence from top to bottom. The white region has been cut out of the layer. The colored region is the part of the layer that is visible from the top.}
	\label{fig:layers}
\end{figure}

We realized that the best way to improve the mathematical construction was to reverse the order of the layers and make an artistic decision to represent the growth of the fractal through disks of large radii that appear to contract from layer to layer to expose more of the detail. This construction led to a much larger proportion of each layer being visible. (See Figure~\ref{fig:layers}.) The parameters were adjusted carefully to balance this ideal visualization with the restrictions imposed by the materials. We needed to make sure that at no point in a laser-cut layer would the wood and acrylic that remains be so thin as to be easily breakable. Notice that the thinnest bridge in version 2 occurs at the northernmost point on the green layer, shown in Figure~\ref{fig:layers}(d), which was a vast improvement over the extremely thin bridge in the red layer of version 1, shown in Figure~\ref{fig:layers-v1}(a).

This was also the first prototype that achieved spacing between the layers of wood by alternating between layers of wood and laser-cut acrylic. It was at that point that we perceived that having a square boundary shape for the layers meant that the light emitted from the LED strip diffused inconsistently across a given layers. We realized that using a circular boundary shape for the layers would give the LED strip a consistent distance to the art, ensuring uniformity and maximizing brightness. Version 2 also had an issue where the light emitted by the LEDs would unintentionally bleed into adjacent layers of acrylic. This significantly detracted from the stroboscopic effect and would need to be remedied by refining the carriage design. Luckily, both of these were addressable via simple updates that were included in the next iteration.

A key, nontrivial, question that came out of our evaluation was ``What is the largest size the art can~be?''. Because the piece requires parts to be tightly integrated with one another, determining how big we could make the art required us to have a strong understanding of what each component of the construction was going to look like and work within our manufacturing constraints. 
To start, we figured out the maximum size the frame could be. We knew that we were designing a piece that was going to be interacted with in an unsupervised environment hundreds of times over the course of a weekend. Users may handle the piece roughly, so we needed to design the parts with robustness in mind. In order to achieve that, many of the internal 3D printed parts were being printed as a single piece. This was particularly important for the frame as it is the main structural component of the piece. This self-imposed requirement meant that we were limited to printing parts that could fit inside the machine. The $x$-$y$ dimensions of the frame were maximized according to the printers used, and that size defined the overall dimensions of the piece.

Once the size of the frame was determined, it was treated as a `ground truth' and much of the layout, sizing, and design details were downstream from that decision, including the size of the art. The layered wood and acrylic pieces occupy most of the $x$-$y$ space in the piece, and the other internal components are stuffed in the spaces between it and the housing. (See Figure~\ref{fig:exploded}.) This resulted in a design with minimal unused space, which is indicative of a well-considered design.

\medskip
\subsection*{Version 3: Vast Improvements in Everything}

Version 3 included a frame which was designed to accommodate different thickness layers in the art stack. We used it as a test platform to try different combinations of wood and acrylic thicknesses. After experimenting with different combinations, we felt that making every layer 1/8 inch thick was optimal.

Version 3 was also the first prototype with handles, which allowed us to replicate the tactile experience of a user for the first time. Although the feeling of using the handles was enchanting, integrating the handle sub-assembly to the carriage was frustrating. The 3D printed coupling attached to the carriage using screws, but the location of the holes with respect to the housing prevented the use of a screwdriver. Resolving this required a re-architecture of the handle sub-assembly. 

This prototype had all of the functionality of the final piece but lacked any cosmetic or user-focused considerations besides the art itself. Knowing that we had most of the basics done, we felt like we were ready to produce a prototype that could replicate `the full experience' we wanted to deliver.

\medskip
\subsection*{Version 4: Refining smaller details}

Version 4 was the first prototype with a wood housing, meaning all of the engineering components were hidden and a viewer of the piece would finally be able to focus on the mathematical visualization. At this point it felt like we had a prototype that was representative of our vision and concluded that there were no critical changes required. We still had the time and energy to iterate once more, so we asked ourselves ``What minor changes can we make toward our guiding principle?''

Because we were interested in making the art reliable and powerful, we wanted to identify ways in which the piece could cause the user to have an unreliable or weak experience. This led us to looking for ways in which the piece could cause a user to focus on something other than the art itself. Every time a user notices something that is imperfect about the piece, their attention migrates away from the art and toward the flaw. These distractions translate to a sub-optimal experience for each individual user, and variance in the experience from one user to the next.

Identifying those distractions relied on our ability to evaluate the piece objectively and nitpick at every flaw we found. As a team composed of one experienced engineer and one experienced mathematician, we found this to be quite natural. Even though the art was captivating, the piece had numerous imperfections, most prominent of which was the surface finish on the wood. The wood was untreated, meaning it was prone to absorbing oils and other liquids it came in contact with. Additionally, the color was too light and made the light from the LEDs appear washed out. Many of the other flaws identified were seemingly minor, and did not impact the art at all. Included in this were things like the weight of the piece, unwanted noise, visible screws, and other annoyances.

\medskip
\subsection*{Version 5: Ready to display}

This was the last version we assembled and the piece shared in the 2024 JMM Mathematical Art Exhibit and is shown in Figure~\ref{fig:fractalemergence}(a). The updates spanned nearly every module of the piece, each of which nudged it closer to our vision. The most impactful update made was the inclusion of three large ceramic tiles. This was done in order to prevent the piece from moving when being played with, and in retrospect was crucial to the piece feeling refined and like it had been produced with craftsmanship. Executing on this involved updating the frame to make space for the tiles and adding a bottom cover to ensure the tiles were properly constrained.

Other refinements were numerous but minor. In particular, we were able to address all of the issues surrounding the surface finish of the wood by adding a spray based polyurethane coating, updated the handle assembly to hide screws, added felt to the bottom cover which allowed the piece to be placed on an non-flat table with wobbling, added a component to help constrain the LED strip to aide in assembly, and added small pieces of foam to prevent the carriage from making noise when it bottomed out.

Even though the piece was shared publicly there were still opportunities for improvement; nothing is ever truly complete. Due to time constraints we chose not to include some of the potential improvements identified in the previous version such as sanding the edges of the laser-cut wood. On top of this, there were other imperfections identified for the first time. Due to a minor assembly issue, the carriage made contact with the housing when being actuated and cause a minor squeaking noise. Luckily, the issue was able to be resolved for the exhibition, but design improvements could prevent the issue from happening in future versions. Additionally, due to an oversight during the final assembly, the LED strip was cut too short, meaning its integrated control panel was inside the frame and not accessible.

\section*{Conclusion}

The iterative development process is invaluable for the construction of physical pieces of mathematical art. The artistic process used to craft the mathematical visualization substantially mirrored the development process used to design the box's mechanisms. Taking the time to thoroughly evaluate each prototype allowed us to move ever closer to the ideal form.  

We perceived another parallel for the process we experienced. A sculptor who is creating a statue from a large block of marble might approach this work in stages by outlining the ideal form, performing a rough cut, then defining features, and finally perfecting details. In our process, we also proceeded through different levels of honing in on the final design by first determining guiding principles, assembling a rough construction of the mechanisms and visualization, making large improvements to both of them, and perfecting the final details. 

The iterative development process integrates powerfully with the technologies that have become ubiquitous in the past twenty years. The ability to work with powerful design software, 3D printers, and laser cutters makes the development of prototypes more efficient which reduces the time between iterations. Furthermore, online marketplaces are able to deliver an unimaginable number of off-the-shelf components within days (and sometimes hours). These components saved time and money, and were likely better quality than what we could have fabricated. It is safe to say that {\em Fractal Emergence} was enabled by these technologies and would have been much more difficult to bring to fruition even 10 years ago. We invite the members of the mathematical art community to explore and evaluate the impact of using these technologies in your own work.

One aspect of how the piece was created is that the layers of wood and acrylic can be removed and replaced by a different scene. Indeed, the authors developed a second piece titled {\em Hyperbolic Emergence}~\cite{Hanusa2} that is based on hyperbolic geometry. One could argue that the frame should be considered a platform that can be used to display many other concepts in an interactive way. 

We aimed to create an interactive piece that gave people the agency to investigate the ideas in their own way and in their own time. The connection between the tactile and visual experience invites the viewer to linger, explore, and ponder. The feedback we received from the attendees conveyed that we achieved this goal with {\em Fractal Emergence}. Readers are encouraged to experience the video available at \cite{Hanusa2}.
    
{\setlength{\baselineskip}{13pt} 
\raggedright				

} 
   
\end{document}